\documentclass[a4paper,fleqn]{cas-dc}

\usepackage[authoryear]{natbib}

\def\tsc#1{\csdef{#1}{\textsc{\lowercase{#1}}\xspace}}
\tsc{WGM}
\tsc{QE}

\begin{document}
\let\WriteBookmarks\relax
\def\floatpagepagefraction{1}
\def\textpagefraction{.001}

\shorttitle{D-Net for Medical Image Segmentation}    

\shortauthors{}  

\title [mode = title]{D-Net: Dynamic Large Kernel with Dynamic Feature Fusion for Volumetric Medical Image Segmentation}  

\author[1]{Jin Yang}
\ead{yang.jin@wustl.edu}

\author[1]{Peijie Qiu}

\author[2]{Yichi Zhang}

\author[1]{Daniel S. Marcus}

\author[1,3]{Aristeidis Sotiras}


\ead{aristeidis.sotiras@wustl.edu}
\cortext[2]{Corresponding author}
\cormark[2]


\affiliation[1]{organization={Mallinckrodt Institute of Radiology},
            addressline={Washington University School of Medicine in St. Louis}, 
            city={St. Louis},
            postcode={63110}, 
            state={MO},
            country={USA}}
            
\affiliation[2]{organization={School of Data Science},
            addressline={Fudan University}, 
            city={Shanghai},
            state={Shanghai},
            country={China}}
            
\affiliation[3]{organization={Institute for Informatics, Data Science and Biostatistics},
            addressline={Washington University School of Medicine in St. Louis}, 
            city={St. Louis},
            postcode={63110}, 
            state={MO},
            country={USA}}

\begin{abstract}
Hierarchical Vision Transformers (ViT) have achieved significant success in medical image segmentation due to their large receptive field and capabilities of leveraging long-range contextual information. Convolutional neural networks (CNNs) may also deliver a large receptive field by using large convolutional kernels. However, due to the employment of fixed-sized kernels, CNNs incorporated with large kernels remain limited in their ability to adaptively capture multi-scale features from organs with large variations in shape and size. They are also unable to utilize global contextual information efficiently. To address these limitations, we propose lightweight Dynamic Large Kernel (DLK) and Dynamic Feature Fusion (DFF) modules. The DLK employs multiple large kernels with varying kernel sizes and dilation rates to capture multi-scale features. Subsequently, a dynamic selection mechanism is utilized to adaptively highlight the most important channel and spatial features based on global information. The DFF is proposed to adaptively fuse multi-scale local feature maps based on their global information. We construct a DLK-Net for medical image segmentation by incorporating DLK and DFF into a hierarchical ViT architecture. Large convolutional kernels are incorporated into hierarchical ViT architectures to utilize their scaling behavior, but they are unable to sufficiently extract low-level features due to feature embedding in ViT architectures. To tackle this limitation, we propose a Salience layer to extract low-level features from images at their original dimensions without feature embedding. This Salience layer employs a Channel Mixer to effectively capture global representations. We incorporated DLK, DFF, and the Salience layer into a hierarchical ViT architecture to develop a novel architecture, termed D-Net. D-Net can effectively utilize a multi-scale large receptive field and adaptively harness global contextual information. To further demonstrate the superiority of our DLK, we incorporated it into a widely used hybrid CNN-ViT architecture to build the DLK-NETR. We apply these three models, including DLK-Net, D-Net, and DLK-NETR, to three volumetric segmentation tasks, and extensive experimental results demonstrate their superior segmentation performance compared to state-of-the-art models, with comparably lower computational complexity.
\end{abstract}


\begin{highlights}
    \item We propose a lightweight \textbf{Dynamic Large Kernel} module for generic feature extraction. DLK employs multiple large convolutional kernels to capture multi-scale features via a large receptive field. It subsequently leverages a dynamic selection mechanism to adaptively highlight important channel and spatial features based on global contextual information.
    \item We propose a lightweight \textbf{Dynamic Feature Fusion} module for adaptive feature fusion. DFF is designed to adaptively fuse multi-scale local features based on global information via dynamic mechanisms.
    \item We propose a \textbf{DLK-Net} to adopt hierarchical transformer behaviors by incorporating DLK and DFF modules into a hierarchical ViT architecture. To demonstrate the superiority of the DLK module, we incorporated the DLK module into a hybrid CNN-ViT architecture to construct a new model, termed \textbf{DLK-NETR}.
    \item We propose a \textbf{Salience layer} for low-level feature extraction and explore its effectiveness. It employs a \textbf{Channel Mixer} for global feature representations.
    \item We design \textbf{D-Net} by incorporating the Salience layer into the DLK-Net. We propose to use these three models, including DLK-Net, DLK-NETR, and D-Net, for 3D volumetric medical image segmentation. They were evaluated on three fundamentally different segmentation tasks and achieved superior segmentation accuracy with comparably lower model complexity over state-of-the-art methods.
\end{highlights}



\maketitle

\section{Introduction}
Segmentation of organs or lesions in medical images is crucial in supporting clinical workflows, including diagnosis, prognosis, and treatment planning. However, manual segmentation is time-consuming and error-prone, thus motivating the development of automatic segmentation tools. Among various tools, deep learning-based methods are widely explored (\cite{chen2022recent}). In recent years, the development of the Vision Transformer (ViT) has led to significant improvements in computer vision tasks (\cite{dosovitskiy2020image}). The key factor in their success is the attention mechanism, which empowers ViT-based models with large receptive fields to utilize global contextual information across the entire input image. However, it faces challenges in serving as a general-purpose backbone due to the high computational complexity of self-attention in high-resolution images. To reduce the complexity of ViT, hierarchical ViTs have been proposed (\cite{liu2021swin,wang2021pyramid,wu2021cvt}). They are more efficient in modeling dense features at various scales, approximating self-attention with a linear complexity. Due to their superior performance, hierarchical ViT-based models have recently been proposed for medical image segmentation and achieved great success (\cite{cao2022swin,peiris2022robust,zhou2023nnformer,qiu2024agileformer,huang2021missformer}). However, the attention mechanism often restricts (hierarchical) ViT-based models from effectively extracting local contextual information. This is because it primarily learns low-resolution features to model global dependencies from 1D sequences, resulting in these features lacking detailed localization information (\cite{chen2021transunet}).

Compared with ViTs, another widely used backbone, Convolutional Neural Networks (CNNs), is more advantageous in local feature extraction (\cite{ronneberger2015u}). However, CNNs are limited in their ability to utilize global contextual information due to the constraints imposed by convolutional kernels (\cite{hu2018squeeze}). To enlarge their receptive fields, convolutional layers with large kernels (LKs) were introduced in CNN architectures (\cite{ding2022scaling,liu2022more,liu2022convnet}). Currently, CNNs with LKs are also attracting attention in medical image segmentation (\cite{azad2024beyond,lee20223d,yang2023ucunet}). However, these networks rely on single fixed-sized large kernels for feature extraction, limiting their ability to capture multi-scale features from organs with large inter-organ and inter-subject variations in shape and size. Additionally, they lack mechanisms to enhance interactions between local features and global contextual information.

To utilize the scaling behavior of hierarchical ViT architectures, convolutional layers with LKs have recently been incorporated into them (\cite{liu2022convnet,woo2023convnext,guo2022segnext}). However, incorporating LK into hierarchical ViT architectures may limit it from sufficiently extracting low-level features. This is because input images are projected to features with lower dimensions by feature embedding in the stem of these ViT architectures. A few models utilize convolutional blocks or residual blocks in the first layer of the decoder to extract low-level features from input images at their original resolution (\cite{li2023large,azad2024beyond,lin2022ds}). However, the effectiveness of utilizing these blocks in extracting low-level features has not been explored. Additionally, employing small convolutional kernels (i.e., $3\times3\times3$) limits models from learning global representations effectively from images with original resolutions (i.e., $96\times96\times96$ and $128\times128\times128$).

To address these limitations, we propose a Dynamic Large Kernel (DLK) module, a Dynamic Feature Fusion (DFF) module, and a Salience layer. In DLK, we propose to use multiple varying-sized large depthwise convolutional kernels. These kernels enable the networks to capture multi-scale contextual information, effectively handling large variations in shape and size. Instead of aggregating these kernels in parallel as in Atrous Spatial Pyramid Pooling (ASPP) (\cite{chen2017deeplab}), we propose to sequentially aggregate multiple large kernels to enlarge receptive fields. Subsequently, we propose a dynamic selection mechanism to adaptively select the most informative local features based on channel-wise and spatial-wise global contextual information. Additionally, we propose the DFF module to fuse multi-scale features based on global information adaptively. During fusion, a channel-wise dynamic selection mechanism is used to preserve the important feature maps, and subsequently, a spatial-wise dynamic selection mechanism is utilized to highlight important spatial regions. We incorporated DLK and DFF modules into a hierarchical ViT architecture to build the DLK-Net for 3D volumetric medical image segmentation. To further demonstrate the superiority of our DLK module, we constructed a DLK-NETR by incorporating the DLK module into a hybrid CNN-ViT architecture.

Subsequently, we propose the Salience layer to extract low-level features from images at their original dimension and explore its effectiveness. In the Salience layer, we employ a Channel Mixer to learn global representations and enhance the feature interaction among channels. We further incorporated the Salience layer into the DLK-Net to develop a novel architecture for 3D volumetric medical image segmentation, termed D-Net. We evaluated these three models, including DLK-Net, DLK-NETR, and D-Net, on three segmentation tasks: abdominal multi-organ segmentation, brain tumor segmentation, and hepatic vessel tumor segmentation. The proposed models outperformed other state-of-the-art models with comparably lower computational complexity. Our contributions are summarized as follows:

\begin{itemize}
    \item We propose a lightweight \textbf{Dynamic Large Kernel} module for generic feature extraction. DLK employs multiple large convolutional kernels to capture multi-scale features via a large receptive field. It subsequently leverages a dynamic selection mechanism to adaptively highlight important channel and spatial features based on global contextual information.
    \item We propose a lightweight \textbf{Dynamic Feature Fusion} module for adaptive feature fusion. DFF is designed to adaptively fuse multi-scale local features based on global information via dynamic mechanisms.
    \item We propose a \textbf{DLK-Net} to adopt hierarchical transformer behaviors by incorporating DLK and DFF modules into a hierarchical ViT architecture. To demonstrate the superiority of the DLK module, we incorporated the DLK module into a hybrid CNN-ViT architecture to construct a new model, termed \textbf{DLK-NETR}.
    \item We propose a \textbf{Salience layer} for low-level feature extraction and explore its effectiveness. It employs a \textbf{Channel Mixer} for global feature representations.
    \item We design \textbf{D-Net} by incorporating the Salience layer into the DLK-Net. We propose to use these three models, including DLK-Net, DLK-NETR, and D-Net, for 3D volumetric medical image segmentation. They were evaluated on three fundamentally different segmentation tasks and achieved superior segmentation accuracy with comparably lower model complexity over state-of-the-art methods.
\end{itemize}

\begin{figure*}[!t]
\centering
\includegraphics[width=\textwidth]{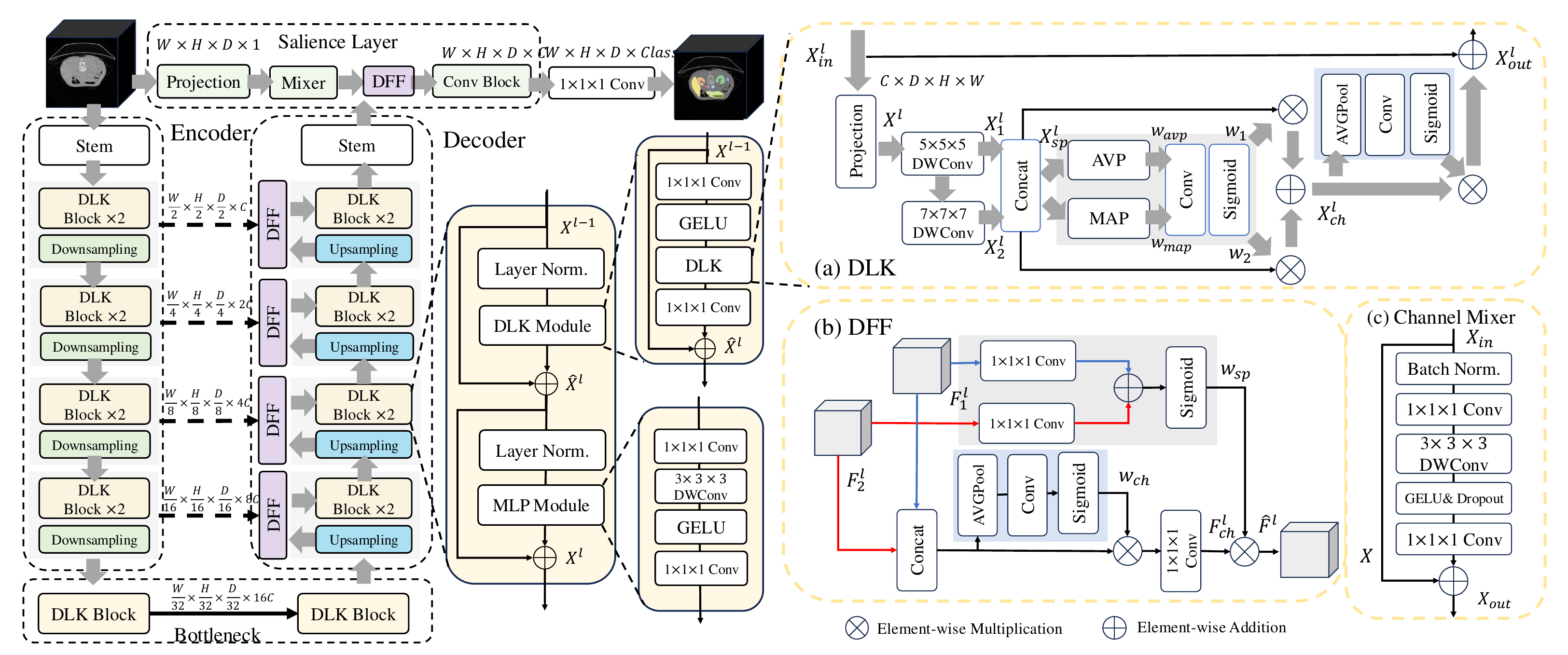}
\caption{The architecture of the D-Net, (a) DLK, (b) DFF, and (c) Channel Mixer. D-Net consists of an encoder, a bottleneck, a decoder, and a Salience layer. Two consecutive DLK blocks are used in each stage for feature extraction. Each DLK block consists of a DLK module and an MLP module. (a) After feature maps $\boldsymbol{X}_{in}^l$ are projected to $\boldsymbol{X}^l$, feature maps $\boldsymbol{X}_1^l$ and $\boldsymbol{X}_2^l$ are extracted by $5\times5\times5$ DWConv and $7\times7\times7$ DWConv, respectively. Subsequently, the dynamic selection values $w_1$ and $w_2$ are generated to calibrate features $\boldsymbol{X}_1^l$ and $\boldsymbol{X}_2^l$. These feature maps $\boldsymbol{X}_{ch}^l$ are scaled based on channel-wise importance. (b) The global channel information $w_{ch}$ is extracted from feature maps $\boldsymbol{F}_1^l$ and $\boldsymbol{F}_2^l$. These feature maps are calibrated to select informative features by a convolution layer as features $\boldsymbol{F}_{ch}^l$. The global spatial information $w_{sp}$ is extracted from $\boldsymbol{F}_1^l$ and $\boldsymbol{F}_2^l$, and is used to recalibrate features $\boldsymbol{F}_{ch}^l$ to generate the adaptively fused features $\hat{\boldsymbol{F}}^l$. (c) Input features $\boldsymbol{X}_{in}$ are mixed across channels to generate output features $\boldsymbol{X}_{out}$.} 
\label{fig1}
\end{figure*}

\section{Related Works}
\subsection{CNN-ViT for Medical Image Segmentation}
ViTs have been combined with CNNs as hybrid networks for medical image segmentation. TransUNet utilized a Transformer to encode tokenized feature maps from CNNs for capturing global information, thus enhancing the segmentation capabilities (\cite{chen2021transunet}). To further enhance its abilities, a Transformer decoder was designed in 3D TransUNet to refine regions of interest adaptively (\cite{chen20233d}). CoTr employed a deformable Transformer to model the long-range dependencies on feature maps from CNNs (\cite{xie2021cotr}). UNETR and Swin UNETR employed a ViT and Swin ViT encoder, respectively, and merged them with the CNN-based decoder for learning feature representations (\cite{hatamizadeh2022unetr,hatamizadeh2021swin}). HiFormer employed a Swin Transformer encoder and a CNN-based encoder to learn global and local feature representations for efficient medical image segmentation (\cite{heidari2023hiformer}).

\subsection{Dynamic Mechanism in Image Tasks}
Various dynamic mechanisms have been applied in CNNs (\cite{han2021dynamic}). Dynamic convolution aggregated multiple convolutional kernels dynamically based on their attention scores (\cite{chen2020dynamic}). A dynamic selection mechanism was proposed to adjust multiple convolutional kernels adaptively based on channel-wise or spatial-wise global information (\cite{li2019selective,li2023large}). Some dynamic mechanisms have also been proposed to improve ViTs. In the Dynamic Window Vision Transformer, a dynamic mechanism was applied to adaptively extract multi-scale features from windows with different sizes (\cite{ren2022beyond}). In the context of medical image segmentation, a dynamic token merging mechanism was utilized to adaptively prepare tokens based on attention scores in DTMFormer (\cite{wang2024dtmformer}).

\section{Method}
\subsection{Dynamic Large Kernel (DLK)}
\textbf{DLK.} We propose the Dynamic Large Kernel (DLK) to adaptively exploit channel-wise and spatial-wise contextual information via a large receptive field (Figure \ref{fig1}a). Specifically, multiple large depthwise kernels are used to extract multi-scale features. Additionally, we cascade these large kernels with growing kernel sizes and increasing dilation rates. This design has two advantages. First, contextual information is aggregated within receptive fields recursively, allowing the effective receptive fields to grow in size progressively (\cite{luo2016understanding}). Second, features extracted within deeper and larger receptive fields contribute more significantly to the output, enabling DLK to capture finer and more informative features. In our work, we utilize a $1\times1\times1$ convolutional layer for channel projection to reduce complexity caused by multiple convolutions, projecting input features $\boldsymbol{X}_{in}^l\in \mathbb{R}^{C\times H \times W\times D}$ to $\boldsymbol{X}^l\in \mathbb{R}^{\frac{C}{2}\times H \times W\times D}$ in the layer $l$ ($C$: the number of channels. $H$, $W$, $D$: the dimension of volumetric images). Subsequently, we use two depthwise convolution (DWConv) layers with large kernels: $\textrm{DWConv}_{(5,1)}$, featuring a $5\times5\times5$ kernel with dilation $1$, and $\textrm{DWConv}_{(7,3)}$, featuring a $7\times7\times7$ kernel with dilation $3$.
\begin{align}
    \nonumber
    \boldsymbol{X}^l & = \textrm{Project}(\boldsymbol{X}_{in}^l)\\
    \nonumber
    \boldsymbol{X}^l_1 &= \textrm{DWConv}_{(5,1)}(\boldsymbol{X}^l)\\
    \nonumber
    \boldsymbol{X}^l_2 &= \textrm{DWConv}_{(7,3)}(\boldsymbol{X}^l_1).
\end{align}
By cascading these kernels, DLK has the same effective receptive field with a $23\times23\times23$ kernel \cite{ding2022scaling}. Features from two LKs, $\boldsymbol{X}^l_1\in \mathbb{R}^{\frac{C}{2}\times H \times W\times D}$ and $\boldsymbol{X}^l_2\in \mathbb{R}^{\frac{C}{2}\times H \times W\times D}$, are concatenated to recover the original number of channels as features $\boldsymbol{X}^l_{sp}\in \mathbb{R}^{C\times H \times W\times D}$. Then the global spatial relationship of these features is efficiently modeled by applying average pooling ($\textrm{AVP}$) and maximum pooling ($\textrm{MAP}$) along channels from features $\boldsymbol{X}^l_{sp}$. 
\begin{align}
    \nonumber
    \boldsymbol{X}^l_{sp} &= \textrm{Concat}([\boldsymbol{X}^l_1;\boldsymbol{X}^l_2]) \\
    \nonumber
    w_{avp} &= \textrm{AVP}(\boldsymbol{X}^l_{sp}) \\
    \nonumber
    w_{map} &= \textrm{MAP}(\boldsymbol{X}^l_{sp}).
\end{align}
Then, a $7\times7\times7$ convolution layer (Conv$_7$) is used to allow such information to interact and mix among different spatial descriptors. Lastly, a Sigmoid activation function is used to obtain dynamic selection values $w_1$, $w_2$.
\begin{align}
\nonumber
    [w_1;w_2] &= \textrm{Sigmoid}(\textrm{Conv}_7([w_{avp};w_{map}])).
\end{align}
Features from different large kernels are adaptively selected by utilizing these selection values to calibrate them.
\begin{align}
    \nonumber
    \boldsymbol{X}^{l}_{ch} &= (w_1 \otimes \boldsymbol{X}^l_{sp}) \oplus (w_2 \otimes \boldsymbol{X}^l_{sp} ).
    \label{eq5}
\end{align} 
Subsequently, the distinctive importance of these features is highlighted by cascading an average pooling (AVGPool), a $1\times1\times1$ convolution layer (Conv$_1$), and a Sigmoid activation. These features $\boldsymbol{X}^l_{ch}\in \mathbb{R}^{C\times H \times W\times D}$ are scaled based on their channel-wise importance. Finally, a residual connection is applied to generate output features $\boldsymbol{X}^l_{out}\in \mathbb{R}^{C\times H \times W\times D}$.
\begin{align}
    \nonumber
    w_{ch}&=\textrm{Sigmoid}(\textrm{Conv}_1(\textrm{AVGPool}(\boldsymbol{X}^l_{ch})))       \\
    \nonumber
    \boldsymbol{X}^l_{out} &= w_{ch} \otimes \boldsymbol{X}^l_{ch} + \boldsymbol{X}^l_{in}.
\end{align}

\textbf{Further Discussions of DLK.} We create a large receptive field in DLK by cascading a depthwise convolutional layer of $5\times5\times5$ kernel with dilation $1$ and a depthwise convolutional layer of $7\times7\times7$ kernel with dilation $3$. It has the effective receptive field (ERF) approximately as the $23\times23\times23$ convolutional kernel. This ERF can be calculated as
\begin{align}
    \nonumber
    R_i=R_{i-1}+(k_i-1)*j_i
\end{align}
where $R_{i-1}$, $k_i$, and $j_i$ are the effective receptive field of the layer $i-1$, kernel size of the layer $i$, and jump of the layer $i$, respectively. To be specific, $R_{i-1}=5$ for the first layer since the first convolutional layer has a kernel size of $5$. $j_i=1$ for the second layer since the stride is equal to $1$. $k_i=19$ for the second layer since the kernel size is $7$ and dilation is $3$. Thus, the ERF of the DLK is $R_i=5+(19-1)*1=23$.

\textbf{DLK module.} The DLK module is built by integrating DLK into two linear layers ($1\times1\times1$ convolution layers; Conv$_1$) with a GELU activation in between. A residual connection is also applied. Accordingly, the output of the $l$-th layer in a DLK module can be computed as
\begin{align}
\nonumber
    \boldsymbol{X}^l&=\textrm{Conv}_1(\boldsymbol{X}^{l-1})\\
    \nonumber
    \boldsymbol{X}^l&=\textrm{DLK}(\textrm{GELU}(\boldsymbol{X}^l))\\
    \nonumber
    \hat{\boldsymbol{X}}^l&=\textrm{Conv}_1(\boldsymbol{X}^l) + \boldsymbol{X}^{l-1}.
\end{align}
\textbf{DLK block.} To leverage the scaling capabilities of hierarchical ViTs, the DLK block is constructed by replacing the multi-head self-attention in a standard hierarchical ViT block with the proposed DLK module. Specifically, the proposed DLK block consists of a DLK module (DLK$^M$) and an MLP module (MLP). Similar to hierarchical ViT blocks, a Layer Normalization (LN) layer is applied before each DLK module and MLP module, and a residual connection is applied after each module. Thus, the DLK block in the $l$-th layer can be computed as
\begin{align}
\nonumber
    \hat{\boldsymbol{X}}^l &= \textrm{DLK$^M$}(\textrm{LN}(\boldsymbol{X}^{l-1}))+\boldsymbol{X}^{l-1}\\
    \nonumber
    \boldsymbol{X}^l &= \textrm{MLP}(\textrm{LN}(\hat{\boldsymbol{X}}^l))+\hat{\boldsymbol{X}}^l.
\end{align}

\subsection{Dynamic Feature Fusion (DFF)}
We propose a Dynamic Feature Fusion (DFF) module to adaptively fuse multi-scale local features based on global information (Figure \ref{fig1}b). It is achieved by dynamically selecting important features based on their global information during fusion. Specifically, feature maps $\boldsymbol{F}_1^l\in \mathbb{R}^{C\times H \times W\times D}$ and $\boldsymbol{F}_2^l\in \mathbb{R}^{C\times H \times W\times D}$ are concatenated along the channel as features $\boldsymbol{F}^l\in \mathbb{R}^{2C\times H \times W\times D}$ 
\begin{align}
    \nonumber
    \boldsymbol{F}^{l}=\textrm{Concat}([\boldsymbol{F}_1^{l};\boldsymbol{F}_2^{l}]).
\end{align}
To ensure the following block can utilize fused features, a channel reduction mechanism is required to reduce the number of channels to the original count $C$. Instead of simply using a $1\times1\times1$ convolution, channel reduction in DFF is guided by global channel information $w_{ch}$. This information is extracted to describe the importance of features by cascading an average pooling (AVGPool), a convolution layer (Conv$_1$), and a Sigmoid activation.
\begin{align}
    \nonumber
    w_{ch}=\textrm{Sigmoid}(\textrm{Conv}_1(\textrm{AVGPool}(\boldsymbol{F}^{l}))).
\end{align}
Fused features are calibrated by the global channel information. Subsequently, a $1\times1\times1$ convolutional layer (Conv$_1$) is utilized to select feature maps based on their importance. This channel information will guide the convolution layer to preserve the important features $\boldsymbol{F}^l_{ch}\in \mathbb{R}^{C\times H \times W\times D}$ while dropping less informative ones.
\begin{align}
    \nonumber
    \boldsymbol{F}_{ch}^l&=\textrm{Conv}_1(w_{ch} \otimes \boldsymbol{F}^{l}).
\end{align}
To model the spatial-wise inter-dependencies among local feature maps, the global spatial information $w_{sp}$ is captured by $1\times1\times1$ convolution layers (Conv$_1$) and a Sigmoid activation from feature maps $\boldsymbol{F}_1^l$ and $\boldsymbol{F}_2^l$. This information is used to calibrate feature maps and put emphasis on salient spatial regions.
\begin{align}
    \nonumber
    w_{sp}&=\textrm{Sigmoid}(\textrm{Conv}_1(\boldsymbol{F}_1^{l})\oplus\textrm{Conv}_1(\boldsymbol{F}_2^{l}))\\
    \nonumber
    \hat{\boldsymbol{F}}^l &=w_{sp} \otimes \boldsymbol{F}_{ch}^l.
\end{align}

\subsection{Salience Layer}
\subsubsection{The Importance of Salience Layer}
Hierarchical ViTs leverage the convolutional stem with a stride of 4 to extract overlapping feature embeddings with the dimension of $\frac{H}{4}\times\frac{W}{4}$ from the input images for image classification (\cite{graham2021levit,xiao2021early}). Classification is an image-level task, and embedding features into lower dimensions improves ViTs to utilize image-level information to make classification with lower computational cost. However, following the same design and utilizing the convolutional stem in the encoder-decoder architecture for segmentation may diminish the performance of ViTs (\cite{xie2021segformer}). Segmentation is a pixel-wise task, and pixel-level information influences the performance of segmentation models. However, the convolutional stem downsamples input images to the dimension of $\frac{H}{4}\times\frac{W}{4}$, leading to loss of pixel-wise information. Thus, ViT-based segmentation models cannot sufficiently utilize pixel-level information from input images and extract low-level features to benefit segmentation. This challenge also limits the performance of some ViT-based models for medical image segmentation (\cite{wang2022mixed,huang2022missformer}).

This challenge has been tackled in CNN-based medical segmentation models, such as U-Net (\cite{ronneberger2015u}) and its variants (\cite{zhou2018unet++,milletari2016v,oktay2018attention}). They employ convolutional blocks to extract low-level features from input images with the original dimension $H\times W$ in the first layer. Additionally, some hybrid CNN-transformer models, including UNETR and its variants (\cite{hatamizadeh2022unetr,hatamizadeh2021swin,shaker2024unetr++}), also tackle this problem by utilizing a convolutional block or a residual block to extract low-level features directly from $H\times W\times D$ input images without downsampling. A ViT-based segmentation network, UNetFormer, solved this problem by employing a transformer block in the decoder to extract features from input images (\cite{hatamizadeh2022unetformer}). This design enhances their segmentation capabilities. Therefore, we are motivated to incorporate a Salience layer to fully utilize pixel-level information and extract low-level features from input images.

We propose that several different blocks can be utilized in the Salience layer, including
\begin{itemize}
    \item ConvBlock. A convolutional block consists of two consecutive $3\times3\times3$ convolutional layers and it was utilized in other methods (\cite{lin2022ds,azad2024beyond}).
    \item DLK. Two consecutive DLKs can be utilized in the Saliency layer to adaptively capture multi-scale low-level features.
    \item Channel Mixer. The performance and efficiency of MLP Mixers in image classification tasks have been described (\cite{tolstikhin2021mlp}), but its potential to extract dense features for medical image segmentation has not been fully explored. We propose to utilize a Channel Mixer in the Saliency Layer to capture low-level features from the original images.
\end{itemize}

\subsubsection{Channel Mixer in Salience Layer}
The core component in the Salience layer is a Channel Mixer. Channel Mixer has a global receptive field which enables it to process the whole visual content. It also allows features to interact among channels. Thus, employing a Channel Mixer in the Salience layer enhances the network to capture global low-level features from input images at their original dimension (Figure \ref{fig1}c). A Batch Normalization (BN) is utilized first. Then, a $1\times1\times1$ convolutional layer (Conv$_1$) is applied to expand channels of $\boldsymbol{X}_{in}\in \mathbb{R}^{C\times H \times W\times D}$ by a ratio of $M=4$ as $\boldsymbol{X}\in \mathbb{R}^{4C\times H \times W\times D}$. The core layer in the Channel Mixer is a $3\times3\times3$ depthwise convolution (DWConv) followed by a GELU activation. Another $1\times1\times1$ convolutional layer (Conv$_1$) is utilized after the GELU to compress channels to the original count as $\boldsymbol{X}_{out}\in \mathbb{R}^{C\times H \times W\times D}$. Two dropout layers are utilized. Lastly, a residual connection is applied to the whole Mixer. This Channel Mixer can be represented as
\begin{align}
    \nonumber
    &\boldsymbol{X}'_{in}=\textrm{BN}(\boldsymbol{X}_{in})\\
    \nonumber
    &\boldsymbol{X}=\textrm{Conv}_1(\boldsymbol{X}'_{in})\\
    \nonumber
    &\boldsymbol{X}=\textrm{Dropout}(\textrm{GELU}(\textrm{DWConv}(\boldsymbol{X})))\\
    \nonumber
    &\boldsymbol{X}_{out}=\textrm{Dropout}(\textrm{Conv}_1(\boldsymbol{X})) + \boldsymbol{X}_{in}.
\end{align}

\begin{figure*}[!t]
\centering
\includegraphics[width=0.7\textwidth]{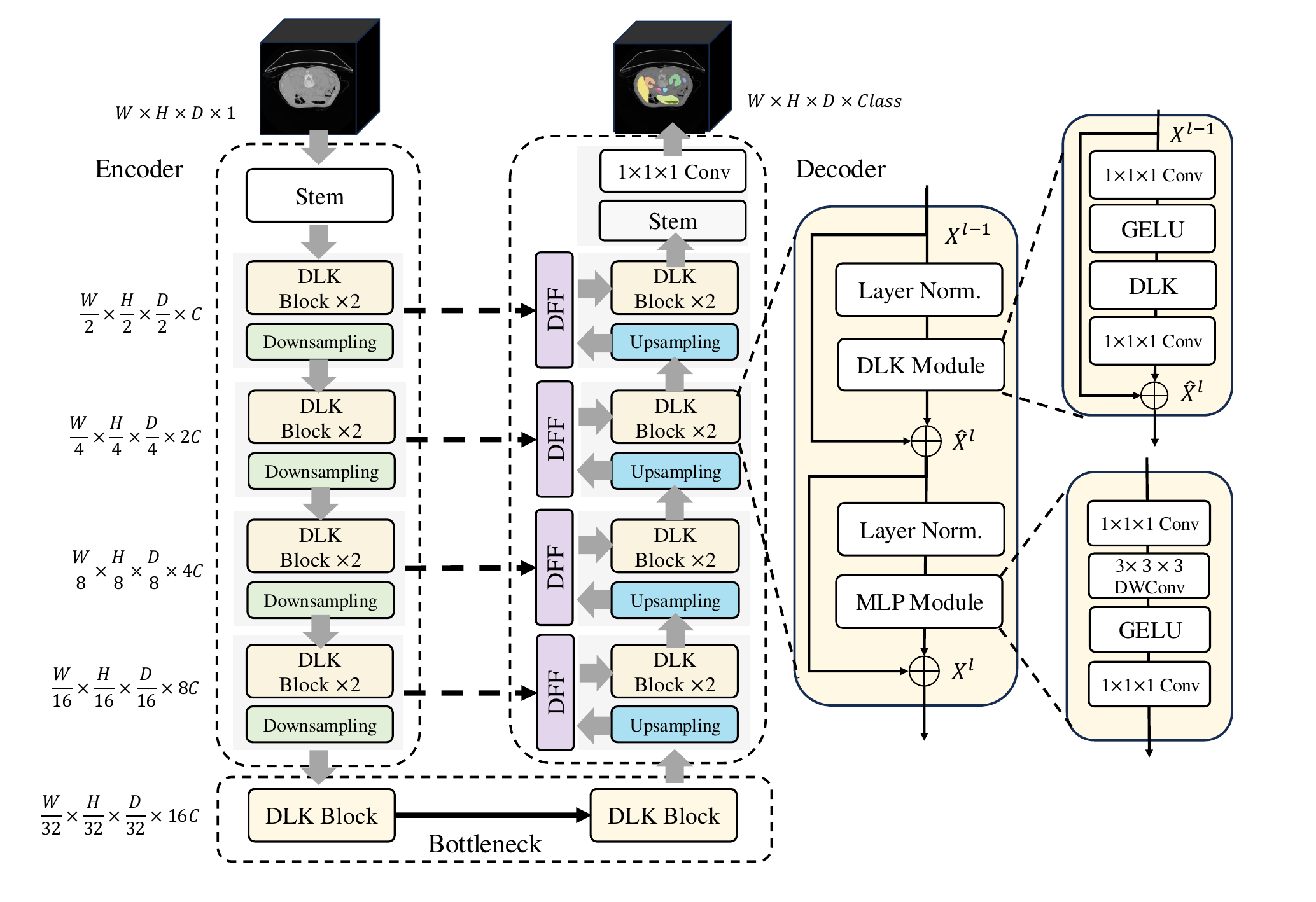}
\caption{The architecture of the DLK-Net. DLK-Net consists of an encoder, a bottleneck, and a decoder. Two consecutive DLK blocks are used in each stage for feature extraction. Each DLK block consists of a DLK module and an MLP module.} 
\label{fig2}
\end{figure*}

\begin{figure*}[!t]
\centering
\includegraphics[width=0.8\textwidth]{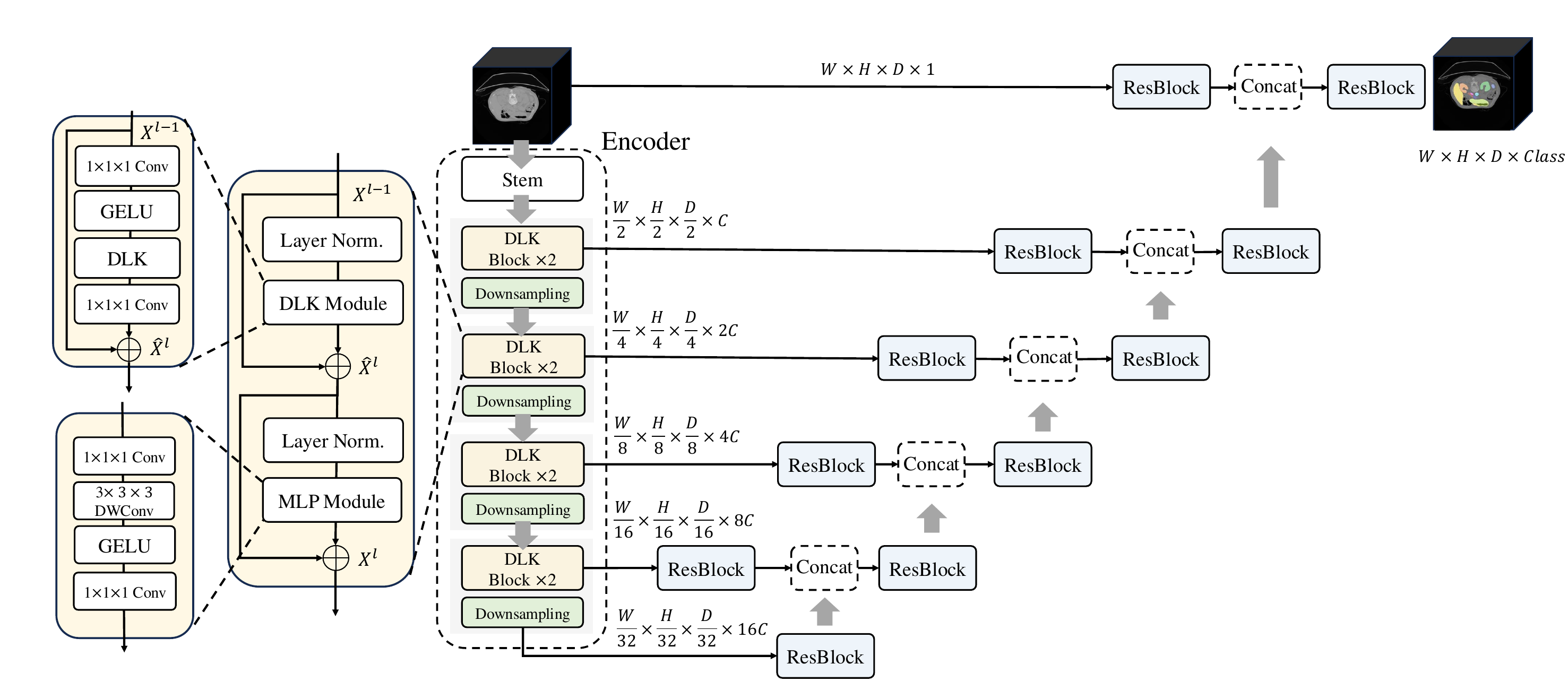}
\caption{The architecture of the DLK-NETR. DLK-NETR utilized the same encoder as DLK-Net and D-Net. In this encoder, two consecutive DLK blocks are used in each stage for feature extraction. Each DLK block consists of a DLK module and an MLP module. This encoder is incorporated into a hybrid CNN-ViT architecture which is the same as UNETR \cite{hatamizadeh2022unetr}, Swin UNETR \cite{hatamizadeh2021swin}, UX Net \cite{lee20223d}, and VSmTrans \cite{liu2024vsmtrans}.} 
\label{fig3}
\end{figure*}

\subsection{Overall Architecture}
The D-Net consists of an encoder, a bottleneck, a decoder, and a Salience layer (Figure \ref{fig1}). The U-shaped encoder-decoder architecture is responsible for learning hierarchical feature representations, and the Salience layer is utilized to extract salient low-level features from original images. These low-level features from the Salience layer are fused with high-level features from hierarchical architecture, thus improving the overall segmentation performance.

\textbf{Encoder.} Instead of flattening the patches and projecting them with linear layers in Hierarchical ViTs, we utilize a large $7\times7\times7$ convolution with a stride of $2$ in the stem to partition the image into feature embeddings with size $\frac{H}{2}\times\frac{W}{2}\times\frac{D}{2}$ and to project them to $C$-dimensional vectors ($C=48$). At each stage, two consecutive DLK blocks are combined to extract contextual information. To exchange the information across channels in the downsampling block, we use a $3\times3\times3$ convolutional layer with a stride of $2$ to downscale the feature map and increase the number of channels by a factor of $2$. Due to the repetitive downsampling, the dimensions of the output feature maps at each stage are $\frac{H}{4}\times\frac{W}{4}\times\frac{D}{4}\times 2C$, $\frac{H}{8}\times\frac{W}{8}\times\frac{D}{8}\times 4C$, $\frac{H}{16}\times\frac{W}{16}\times\frac{D}{16}\times 8C$, and $\frac{H}{32}\times\frac{W}{32}\times\frac{D}{32}\times 16C$, respectively.

\textbf{Bottleneck.} Two consecutive DLK blocks are used for the bottleneck to extract high-level features. The dimensions of both input and output features are $\frac{H}{32}\times\frac{W}{32}\times\frac{D}{32}\times 16C$.

\textbf{Decoder.} At each stage, a $2\times2\times2$ transposed convolutional layer with a stride of $2$ is used to upscale the feature map and decrease the number of channels by both a factor of 2. These upsampled features are then fused with the features from the encoder via skip connections within DFF modules. Subsequently, two consecutive DLK blocks are used. The dimensions of the output feature maps at each stage are $\frac{H}{16}\times\frac{W}{16}\times\frac{D}{16}\times 8C$, $\frac{H}{8}\times\frac{W}{8}\times\frac{D}{8}\times 4C$, $\frac{H}{4}\times\frac{W}{4}\times\frac{D}{4}\times 2C$, and $\frac{H}{2}\times\frac{W}{2}\times\frac{D}{2}\times C$, respectively. In the stem of the decoder, a $2\times2\times2$ transposed convolutional layer with a stride of $2$ is used to upsample the feature maps to a dimension of $H\times W\times D\times C$.

\textbf{Salience layer.} A channel projection block is utilized to project the number of input channels to $C$ without changing the resolution of the input image. It consists of a $3\times3\times3$ convolutional layer, a Batch Normalization, and a Leaky ReLU. Then a Channel Mixer is employed to extract low-level spatial features and aggregate features across channels. These features are fused with those from the decoder within a DFF module. Subsequently, a convolution block, consisting of two consecutive $3\times3\times3$ convolutional layers, is used to capture finer features. Lastly, a $1\times1\times1$ convolutional layer is used to produce the voxel-wise segmentation prediction.

\subsection{Model Variants}
\textbf{DLK-Net.} We propose a lightweight variant of the D-Net for efficient medical image segmentation, termed DLK-Net (Figure \ref{fig2}). It is constructed by incorporating the DLK and DFF modules into a hierarchical ViT architecture, and the Salience layer is not utilized in the DLK-Net. To be specific, it employs a symmetric U-shaped ViT architecture, consisting of the encoder, the bottleneck, and the decoder of the D-Net.

\textbf{DLK-NETR.} We propose another variant of the D-Net by incorporating the DLK module into a hybrid CNN-ViT architecture, termed DLK-NETR (Figure \ref{fig3}). Several volumetric segmentation methods, such as UNETR (\cite{hatamizadeh2022unetr}), Swin UNETR (\cite{hatamizadeh2021swin}), UX Net (\cite{lee20223d}), and VSmTrans (\cite{liu2024vsmtrans}), employ the same hybrid CNN-ViT architecture. The difference between them is the employment of different modules in the encoder. Specifically, UNETR and Swin UNETR employ the basic Transformer module and Swin Transformer module, respectively, while UX Net employs an LK module. In contrast, VSmTrans employs a module that tightly integrates self-attentions and convolutions. Thus, we are motivated to utilize the DLK module in the encoder, and then incorporate this encoder into the same hybrid architecture to generate DLK-NETR. We compare our DLK-NETR with other models to demonstrate the superiority of the DLK module.

\section{Experiments and Results}
\subsection{Datasets} 
\textbf{Abdominal Multi-organ Segmentation.} We conducted experiments on the MICCAI 2022 AMOS Challenge dataset (AMOS 2022) for abdominal multi-organ segmentation in Computed Tomography (CT) images (\cite{ji2022amos}). It consists of 300 multi-contrast abdominal CT images, and each image has 15 anatomical organs with manual annotations. Anatomical labels include Spleen, Right kidney, Left kidney, Gall bladder, Esophagus, Liver, Stomach, Arota, Postcava, Pancreas, Right Adrenal Gland, Left Adrenal Gland, Duodenum, Bladder, and Prostate. 3D volumes were pre-processed to volumetric patches with a dimension of $96\times 160 \times 160$. 5-fold cross-validation was applied and each fold includes 60 scans.

\textbf{Brain Tumor Segmentation.} Brain Tumors Segmentation dataset is from the Medical Segmentation Decathlon (MSD) Challenge (\cite{antonelli2022medical}). It consists of 484 multi-parametric Magnetic Resonance Imaging (MRI) scans with segmentation labels. Four modalities are available for each participant: Native T1-weighted image (T1w), post-contrast T1-weighted (T1Gd), T2-weighted (T2w), and T2 Fluid Attenuated Inversion Recovery (T2-FLAIR). Each subject has three foreground annotations: Edema (ED), Enhancing Tumor (ET), and Non-Enhancing Tumor (NET). Data was pre-processed to volumetric patches with a dimension of $128\times128\times128$. 5-fold cross-validation was applied and these folds include 97, 97, 97, 97, and 96 scans, respectively.

\textbf{Hepatic Vessel Segmentation.} The Hepatic Vessel Tumor segmentation dataset is from the MSD Challenge (\cite{antonelli2022medical}). It consists of 303 CT scans with manual annotations. The target segmentation regions are the hepatic vessels (Vessel) and tumors within the liver (Tumor). They are obtained from patients with a variety of primary and metastatic liver tumors. 3D volumes were pre-processed to volumetric patches with a dimension of $64\times192\times192$. 5-fold cross-validation was applied and these folds include 61, 61, 61, 60, and 60 scans, respectively.

\subsection{Implementation Details}
The D-Net was implemented using PyTorch\footnote{http://pytorch.org/}. A combination of dice loss $\mathcal{L}_{Dice}$ and cross-entropy loss $\mathcal{L}_{CE}$ was used as the loss function. The loss function $\mathcal{L}$ can be formulated between the prediction $\hat{y}$ and the ground truth $y$ as:
\begin{align}
    \nonumber
    \mathcal{L}=\lambda_1\mathcal{L}_{Dice}(\hat{y},y)+\lambda_2\mathcal{L}_{CE}(\hat{y},y).
\end{align}
where $\lambda_1$ and $\lambda_2$ are weighting hyperparameters and both set as $0.5$. The Stochastic Gradient Descent (SGD) was used as the optimizer. The initial learning rate was set to 0.001 for multi-organ segmentation and hepatic vessel segmentation, and it was set to 0.0001 for brain tumor segmentation. Then the learning rate was decayed with a poly learning rate scheduler during training. Models were trained for 1000 epochs with a batch size of 2 on NVIDIA Tesla A100 PCI-E Passive Single GPU with 40GB of GDDR5 memory. The code is made available at https://github.com/sotiraslab/DLK.

We applied the normalization strategy for CT scans from the AMOS 2022 and MSD Hepatic Vessel dataset and MRI scans from the MSD Brain Tumor dataset. The image intensities were clipped by $5\%$ and $95\%$ of intensity values, and then z-score normalization was applied to each volume. Subsequently, scans were cropped to sub-volumes as input patches with a specific size. Data augmentation techniques were implemented to improve model robustness. To be specific, patches were rotated between $[-30, 30]$ along three axes with a probability of $0.2$ and then scaled between $(0.7,1.4)$ with a probability of $0.2$. Subsequently, all patches were mirrored along all axes with a probability of $0.5$. Zero-centered additive Gaussian noise with the variance drawn from the distribution $U(0, 0.1)$ and brightness was added to each voxel sample with a probability of $0.15$, separately.

The segmentation performance was evaluated using the Dice coefficient score (Dice) and Intersection over Union (IoU). The computational complexity of architectures was evaluated by the number of parameters (Params) and the number of Floating Point Operations (FLOPs).

\begin{table*}[!t]
\centering
\caption{Comparison of segmentation performance among D-Net, DLK-Net, DLK-NETR, and other SOTA methods on the 2022 AMOS Abdominal Multi-organ segmentation task. \textbf{Bold} represents the best results, and \underline{underline} represents the second best results. The overall segmentation performance was evaluated using the Dice and IoU. The organ-specific segmentation performance was evaluated using the Dice (Mean $\pm$ Standard Deviation (SD)). ($^*$: $p<0.01$ with Wilcoxon signed-rank test between D-Net and each SOTA method. $^\dagger$: $p<0.01$ with Wilcoxon signed-rank test between DLK-NETR and each baseline method.)}
\label{tab1}
\resizebox{\textwidth}{!}{
\begin{tabular}{c|c|c|c|c|c|c|c|c|c|c|c|c|c|c}
\toprule
Tasks &  VNet & nnU-Net & Att U-Net & TransBTS & UNETR & nnFormer & Swin UNETR & UX Net & MedNext & SegFormer & VSmTrans & DLK-Net & DLK-NETR & D-Net\\
\midrule
Spleen       & 95.06$\pm$8.44 & 96.37$\pm$6.14 & 96.28$\pm$6.82 & 95.45$\pm$8.35 & 90.09$\pm$11.95 & 94.20$\pm$4.23 & 95.32$\pm$8.53 &
               95.44$\pm$8.91 & 95.63$\pm$8.40 & 92.07$\pm$9.97 & 95.90$\pm$8.33 & 96.59$\pm$2.99 & \underline{96.68}$\pm$1.85 & \textbf{97.60}$\pm$1.48 \\
R. kidney    & 94.74$\pm$8.22 & 95.70$\pm$5.94 & 95.79$\pm$5.98 & 95.35$\pm$5.04 & 89.58$\pm$12.61 & 94.13$\pm$4.23 & 95.00$\pm$8.30 & 
               95.14$\pm$6.82 & 93.79$\pm$9.05 & 92.57$\pm$6.71 & 95.54$\pm$6.17 & 95.98$\pm$3.08 & \underline{96.26}$\pm$1.59 & \textbf{97.06}$\pm$1.69 \\
L. kidney    & 94.84$\pm$8.57 & \underline{95.34}$\pm$9.02 & 95.33$\pm$9.54 & 94.57$\pm$10.95 & 89.05$\pm$13.39 & 93.54$\pm$9.69 & 94.80$\pm$9.92 & 
               94.58$\pm$10.79 & 93.52$\pm$10.97 & 91.63$\pm$11.51 & 94.76$\pm$10.64 & 94.97$\pm$11.04 & 94.39$\pm$11.85 & \textbf{96.60}$\pm$12.48 \\
Gall bladder & 76.80$\pm$25.93 & 82.21$\pm$23.31 & \underline{83.20}$\pm$22.45 & 79.15$\pm$24.35 & 61.78$\pm$28.02 & 77.31$\pm$23.38 & 
               78.08$\pm$25.73 & 78.32$\pm$26.15 & 78.58$\pm$25.61 & 69.84$\pm$25.69 & 81.13$\pm$23.89 & 81.72$\pm$24.29 & 81.47$\pm$24.18 & \textbf{85.01}$\pm$ 21.62 \\
Esophagus    & 80.16$\pm$10.41 & 83.60$\pm$9.97 & 83.87$\pm$9.64 & 80.74$\pm$10.48 & 68.38$\pm$15.32 & 74.27$\pm$13.17 & 81.08$\pm$10.26 & 
               81.80$\pm$10.16 & 80.61$\pm$11.86 & 69.23$\pm$12.71 & 83.22$\pm$9.22 & \underline{84.57}$\pm$7.51 & 83.03$\pm$9.03 & \textbf{86.45}$\pm$ 8.41 \\
Liver        & 96.73$\pm$2.31 & \underline{97.35}$\pm$2.17 & 97.34$\pm$2.40 & 96.96$\pm$1.98 & 93.70$\pm$5.18 & 96.38$\pm$2.59 & 96.81$\pm$2.59 & 
               96.89$\pm$2.95 & 97.02$\pm$2.42 & 95.44$\pm$3.29 & 97.32$\pm$2.01 & 96.92$\pm$3.92 & 97.24$\pm$2.05 & \textbf{97.59}$\pm$1.83 \\
Stomach      & 87.42$\pm$16.30 & 90.34$\pm$15.49 & 90.19$\pm$15.75 & 88.10$\pm$16.89 & 74.31$\pm$19.39 & 86.81$\pm$15.91 & 87.19$\pm$17.03 & 
               87.50$\pm$16.88 & 88.51$\pm$16.42 & 82.40$\pm$17.29 & 89.80$\pm$16.21 & \underline{91.48}$\pm$9.24 & 89.21$\pm$18.37 & \textbf{92.39}$\pm$18.40 \\
Arota        & 91.24$\pm$6.46 & 93.62$\pm$5.32 & 94.31$\pm$4.44 & 92.45$\pm$6.14 & 87.46$\pm$6.58 & 90.45$\pm$6.43 & 93.25$\pm$5.36 & 
               93.26$\pm$5.38 & 89.22$\pm$8.22 & 90.59$\pm$3.72 & \underline{94.94}$\pm$3.22 & 94.57$\pm$4.24 & 94.91$\pm$2.77 & \textbf{96.08}$\pm$2.19 \\
Postcava     & 85.64$\pm$8.75 & 89.37$\pm$6.33 & 89.71$\pm$6.06 & 87.50$\pm$7.40 & 77.61$\pm$9.08 & 82.43$\pm$11.40 & 87.80$\pm$7.03 & 
               88.16$\pm$6.84 & 86.02$\pm$9.56 & 83.34$\pm$6.91 & 90.03$\pm$5.56 & \underline{90.62}$\pm$5.51 & 90.04$\pm$6.01 & \textbf{91.88}$\pm$5.75 \\
Pancreas     & 80.83$\pm$12.27 & 84.43$\pm$11.79 & 84.53$\pm$11.75 & 81.00$\pm$13.94 & 67.45$\pm$16.53 & 76.05$\pm$14.69 & 81.45$\pm$13.47 & 
               82.09$\pm$12.98 & 81.36$\pm$13.98 & 74.67$\pm$13.84 & 84.07$\pm$12.51 & 82.78$\pm$14.40 & \underline{85.47}$\pm$8.62 & \textbf{86.75}$\pm$7.96\\
R.A. gland   & 72.84$\pm$13.12 & 75.71$\pm$12.82 & 75.75$\pm$13.52 & 72.24$\pm$14.10 & 63.09$\pm$14.61 & 64.91$\pm$12.17 & 73.96$\pm$12.72 & 
               74.29$\pm$13.30 & 73.90$\pm$14.02 & 58.66$\pm$11.43 & 75.50$\pm$11.96 & 75.50$\pm$14.07 & \underline{77.20}$\pm$7.95 & \textbf{78.40}$\pm$8.53\\
L.A. gland   & 73.75$\pm$14.64 & 76.40$\pm$14.14 & 76.37$\pm$14.27 & 73.09$\pm$14.26 & 58.91$\pm$18.81 & 64.51$\pm$12.97 & 73.16$\pm$16.02 & 
               74.49$\pm$15.10 & 73.99$\pm$15.26 & 55.01$\pm$13.53 & 75.87$\pm$14.06 & 73.12$\pm$19.27 & \underline{78.48}$\pm$9.09 & \textbf{79.33}$\pm$7.61\\
Duodenum     & 74.59$\pm$15.08 & 79.64$\pm$14.44 & 79.44$\pm$14.55 & 75.55$\pm$15.48 & 60.35$\pm$14.81 & 69.42$\pm$15.33 & 75.44$\pm$14.90 & 
               75.89$\pm$15.21 & 75.45$\pm$16.11 & 67.64$\pm$14.08 & 79.60$\pm$14.09 & 78.91$\pm$16.47 & \underline{80.54}$\pm$11.36 & \textbf{82.73}$\pm$11.47\\
Bladder      & 84.46$\pm$17.78 & 87.90$\pm$15.06 & 88.09$\pm$15.44 & 86.02$\pm$16.49 & 70.17$\pm$22.87 & 82.67$\pm$18.06 & 84.68$\pm$17.84 & 
               85.03$\pm$18.43 & 86.00$\pm$16.71 & 77.80$\pm$18.71 & 87.00$\pm$16.43  & \underline{89.57}$\pm$11.66 & 88.01$\pm$13.49 & \textbf{90.71}$\pm$11.00\\
Prostate     & 78.82$\pm$19.34 & 82.94$\pm$18.78 & 83.18$\pm$18.84 & 81.47$\pm$18.53 & 68.95$\pm$21.83 & 77.58$\pm$18.70 & 79.22$\pm$20.52 & 
               79.90$\pm$21.26 & 80.13$\pm$20.07 & 74.45$\pm$19.26 & 81.55$\pm$19.91 & \underline{84.54}$\pm$16.57 & 81.83$\pm$19.86 & \textbf{86.45}$\pm$17.90\\
\midrule
Mean Dice         & 84.53 & 87.39 & 87.56 & 85.31 & 74.73 & 81.65 & 85.15 & 85.52 & 84.91 & 78.36 & 87.08 & 87.46 & \underline{87.65}$^\dagger$ & \textbf{89.67}$^*$ \\
SD Dice           & 15.98 & 14.47 & 14.48 & 15.78 & 20.31 & 17.11 & 16.17 & 16.15 & 16.19 & 18.57 & 14.91 & 14.74 & 13.65 & 12.56 \\
Mean IoU          & 75.78 & 79.81 & 80.08 & 76.93 & 63.21 & 71.88 & 76.77 & 77.33 & 76.43 & 67.64 & 79.42 & 79.98 & \underline{80.10}$^\dagger$ & \textbf{81.54}$^*$ \\
\bottomrule
\end{tabular}}
\end{table*}

\begin{table*}[!t]
\centering
\caption{Comparison of segmentation performance among D-Net, DLK-Net, DLK-NETR, and other SOTA methods on the MSD Multi-modality Brain Tumor segmentation task. \textbf{Bold} represents the best results, and \underline{underline} represents the second best results.  The overall segmentation performance was evaluated using the Dice and IoU. The organ-specific segmentation performance was evaluated using the Dice (Mean $\pm$ Standard Deviation (SD)). ($^*$: $p<0.01$ with Wilcoxon signed-rank test between D-Net and each SOTA method. $^\dagger$: $p<0.01$ with Wilcoxon signed-rank test between DLK-NETR and each baseline method.)}
\label{tab2}
\resizebox{\textwidth}{!}{
\begin{tabular}{c|c|c|c|c|c|c|c|c|c|c|c|c|c|c}
\toprule
Tasks & VNet & nnU-Net & Att U-Net & TransBTS & UNETR & nnFormer & Swin UNETR & UX Net & MedNext & SegFormer & VSmTrans & DLK-Net & DLK-NETR & D-Net\\
\midrule
ET   & 77.38$\pm$22.46 & 79.27$\pm$21.01 & 78.94$\pm$21.53 & 78.77$\pm$21.57 & 78.29$\pm$21.66 & 79.50$\pm$22.01 & 78.81$\pm$21.92 &
       61.75$\pm$26.36 & 79.04$\pm$21.74 & 40.65$\pm$26.61 & 79.28$\pm$21.19 & 79.31$\pm$20.66 & \underline{79.95}$\pm$20.52 & $\boldsymbol{80.02}$$\pm$20.44 \\
ED   & 79.91$\pm$12.48 & 80.33$\pm$12.13 & 80.41$\pm$11.93 & 80.71$\pm$11.21 & 79.36$\pm$12.10 & \underline{80.99}$\pm$13.50 & 80.80$\pm$11.77 & 
       61.83$\pm$15.20 & 80.71$\pm$11.71 & 64.04$\pm$17.08 & 80.63$\pm$12.06 & 80.18$\pm$12.17 & 80.44$\pm$10.97 & $\boldsymbol{81.08}$$\pm$10.38  \\
NET  & 61.60$\pm$23.10 & 61.88$\pm$23.07 & 61.47$\pm$23.20 & 61.65$\pm$23.14 & 59.74$\pm$23.22 & 61.32$\pm$24.24 & 61.72$\pm$23.45 &
       48.01$\pm$23.51 & 61.97$\pm$22.89 & 34.92$\pm$24.13 & 62.10$\pm$22.94 & 61.43$\pm$23.36 & \underline{62.14}$\pm$21.51 & \textbf{62.16}$\pm$21.58 \\
\midrule
Mean Dice & 72.96 & 73.83 & 73.61 & 73.71 & 72.46 & 73.93 & 73.78 & 57.20 & 73.91 & 46.54 & 74.00 & 73.64 & \underline{74.18}$^\dagger$ & \textbf{74.42}$^*$ \\
SD Dice   & 21.52 & 21.10 & 21.34 & 21.18 & 21.58 & 21.87 & 21.51 & 23.12 & 21.20 & 26.19 & 21.09 & 21.17 & 19.90 & 19.57 \\
Mean IoU  & 61.18 & 62.17 & 61.97 & 62.04 & 60.58 & 62.32 & 62.22 & 43.42 & 62.31 & 34.07 & 62.40 & 62.27 & \underline{62.53}$^\dagger$ & \textbf{62.87}$^*$ \\
\bottomrule
\end{tabular}}
\end{table*}

\begin{table*}
  \centering
  \caption{Comparison of segmentation performance among D-Net, DLK-Net, DLK-NETR, and other SOTA methods on the MSD Hepatic Vessel Tumor segmentation task. \textbf{Bold} represents the best results, and \underline{underline} represents the second best results. The overall segmentation performance was evaluated using the Dice and IoU. The organ-specific segmentation performance was evaluated using the Dice (Mean $\pm$ Standard Deviation (SD)). ($^*$: $p<0.01$ with Wilcoxon signed-rank test between D-Net and each baseline method. $^\dagger$: $p<0.01$ with Wilcoxon signed-rank test between DLK-NETR and each baseline method.)}
  \begin{tabular}{c|ccc|cc}
    \toprule
    Methods   & Mean Dice & SD Dice & Mean IoU & Vessel & Tumor \\
    \midrule
    VNet        & 65.12 & 21.34 & 51.52 & 62.89$\pm$12.37 & 67.35$\pm$27.35 \\
    nnU-Net     & 65.99 & 21.32 & 52.42 & 62.55$\pm$12.84 & 69.43$\pm$21.32 \\
    Att U-Net   & 65.25 & 20.87 & 51.58 & 62.17$\pm$12.53 & 68.33$\pm$26.36 \\
    TransBTS    & 64.28 & 21.25 & 50.57 & 61.60$\pm$12.01 & 66.97$\pm$27.29 \\
    UNETR       & 55.97 & 24.46 & 42.48 & 61.96$\pm$11.73 & 49.98$\pm$31.43 \\
    nnFormer    & 65.23 & 20.68 & 51.47 & 62.62$\pm$11.74 & 67.85$\pm$26.53 \\
    Swin UNETR  & 63.20 & 21.87 & 49.48 & 62.47$\pm$12.38 & 63.92$\pm$28.33 \\
    UX Net      & 64.25 & 20.68 & 50.36 & 62.87$\pm$12.27 & 65.63$\pm$26.47 \\
    MedNext     & 66.16 & 20.09 & 52.39 & 62.57$\pm$12.21 & \underline{69.74}$\pm$25.15 \\
    SegFormer   & 58.82 & 21.75 & 44.83 & 54.42$\pm$10.37 & 63.23$\pm$28.28 \\
    VSmTrans    & 65.01 & 20.20 & 51.13 & 62.05$\pm$12.63 & 67.97$\pm$25.28 \\
    \midrule
    DLK-Net     & 65.61 & 21.05& 51.73 & 61.66$\pm$12.56 & 69.56$\pm$26.55 \\ 
    DLK-NETR    & $\boldsymbol{67.75}^\dagger$ & 20.34 & $\boldsymbol{54.30}^\dagger$ & $\boldsymbol{65.76}$$\pm$10.49 & 69.73$\pm$26.64 \\
    D-Net       & \underline{67.63}$^*$ & 19.34 & \underline{53.59}$^*$ & \underline{65.08}$\pm$11.45 & $\boldsymbol{70.17}$$\pm$24.21   \\
    \bottomrule
  \end{tabular}
  \label{tab3}
\end{table*}

\begin{table}[!t]
\centering
\caption{Comparison of model complexity among D-Net, DLK-Net, DLK-NETR, and other SOTA methods. The number of parameters (Params) and FLOPs are evaluated for input patches with the dimension $96\times96\times96$.}\label{tab4}
\begin{tabular}{c|cc}
\toprule
Methods    & Params (M) & FLOPs (G) \\
\midrule
VNet       & 45.66      & 370.52  \\
nnU-Net    & 68.38      & 357.13  \\
Att U-Net  & 69.08      & 360.98  \\
TransBTS   & 31.58      & 110.69  \\
UNETR      & 92.78      & 82.73   \\
nnFormer   & 149.33     & 284.28  \\
Swin UNETR & 62.19      & 329.28  \\
UX Net     & 53.01      & 632.33  \\
MedNext    & 11.65      & 178.05  \\
SegFormer  & 4.50       & 5.02    \\
VSmTrans   & 50.39      & 358.21  \\
\midrule
DLK-Net    & 39.12      & 62.37   \\
DLK-NETR   & 45.89      & 336.75  \\
D-Net      & 39.28      & 200.13  \\
\bottomrule
\end{tabular}
\end{table}

\begin{figure*}[!t]
\centering
\includegraphics[width=\textwidth]{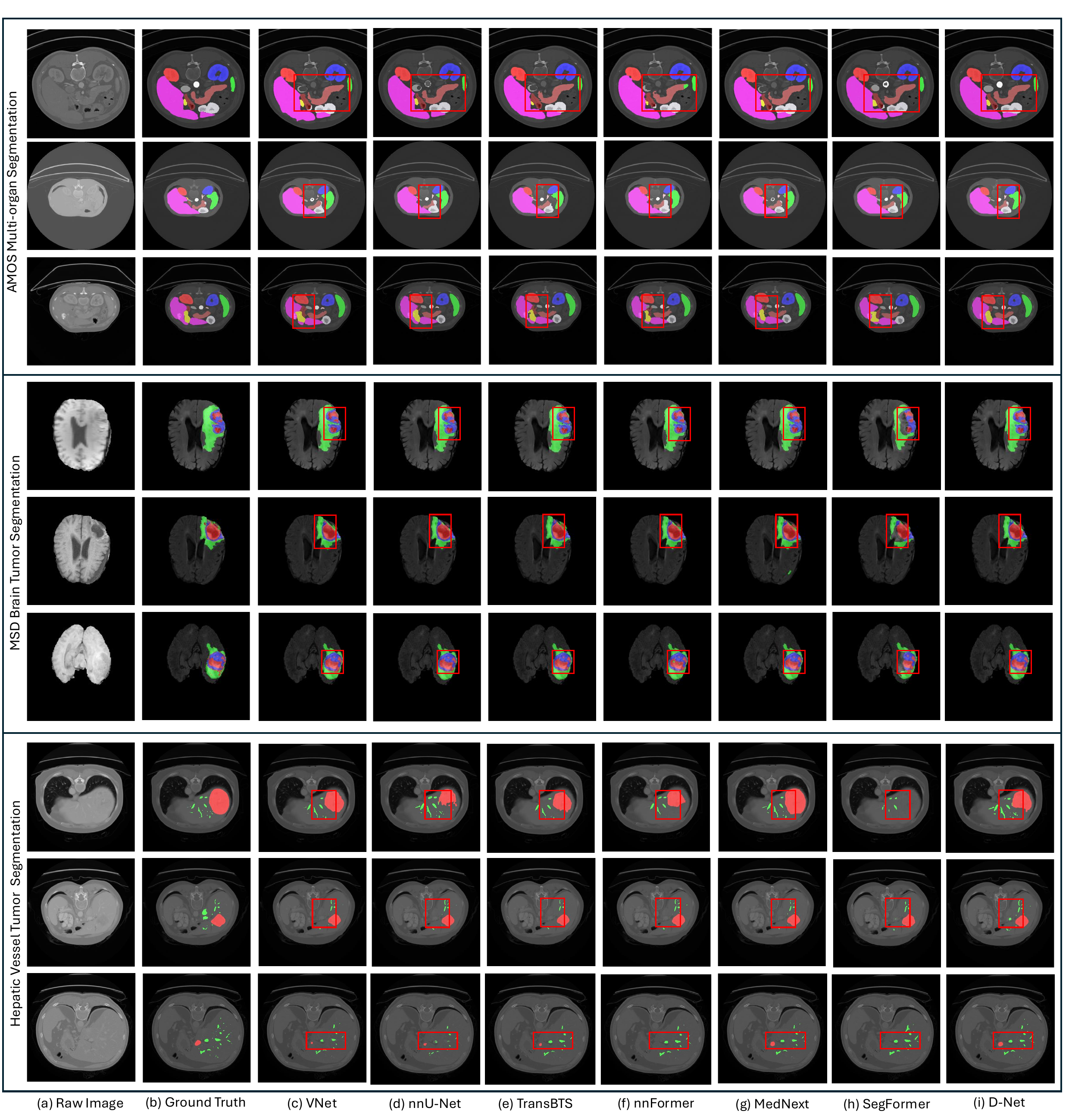}
\caption{Qualitative comparison between D-Net and other methods ((c) VNet (d) nnU-Net (e) TransBTS (f) nnFormer (g) MedNext (f) SegFormer) across three public datasets, including the AMOS 2022 Multi-organ dataset, the MSD Brain Tumor dataset, and the MSD Hepatic Vessel Tumor dataset.} 
\label{vis1}
\end{figure*}

\begin{figure*}[!t]
\centering
\includegraphics[width=\textwidth]{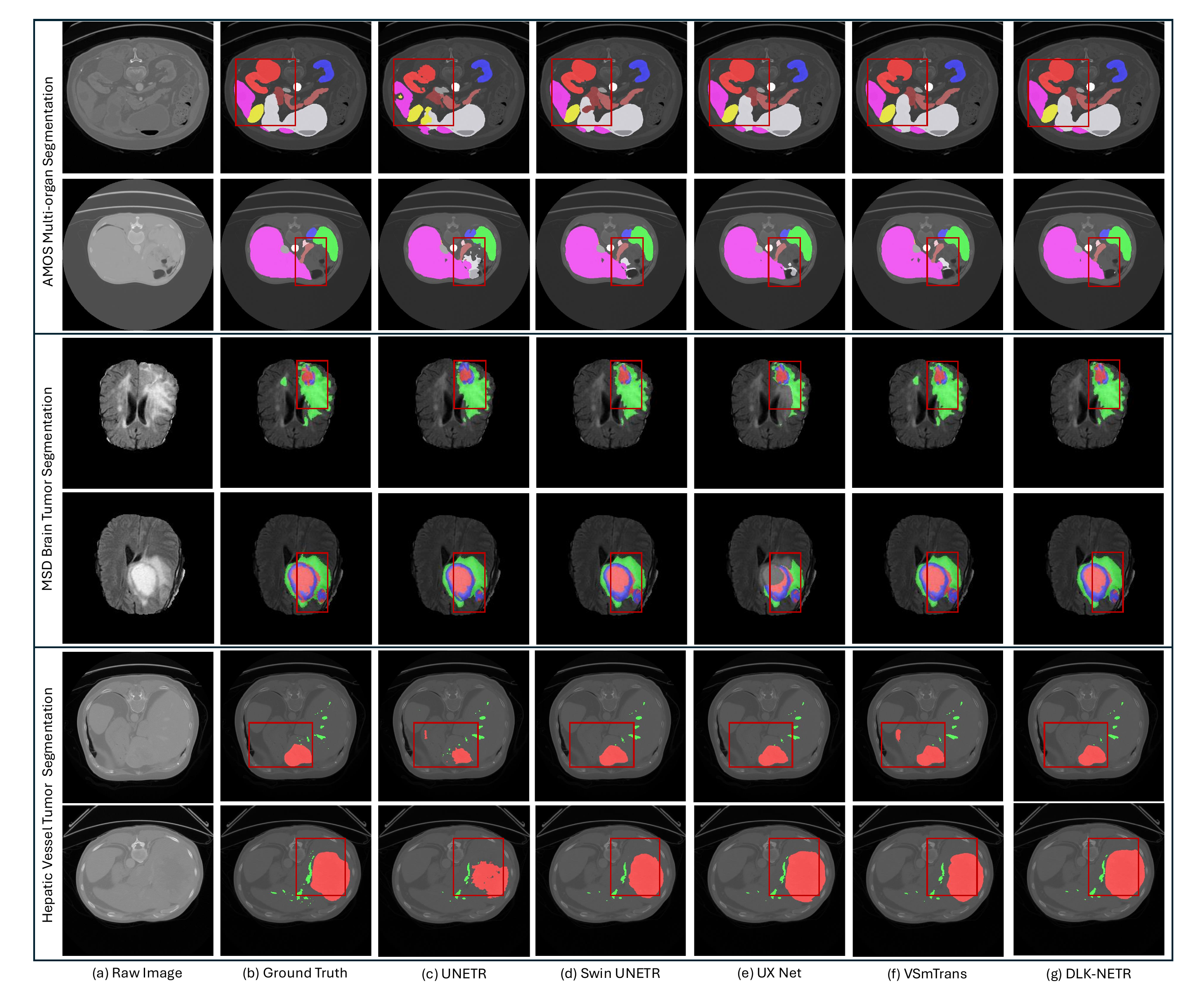}
\caption{Qualitative comparison between DLK-NETR and other methods which employ the same architecture ((c) UNETR (d) Swin UNETR (e) UX Net (f) VSmTrans) across three public datasets, including the AMOS 2022 Multi-organ dataset, the MSD Brain Tumor dataset, and the MSD Hepatic Vessel Tumor dataset.} 
\label{vis2}
\end{figure*}

\subsection{Comparison with State-of-the-arts}
We evaluate the D-Net, DLK-Net, and DLK-NETR on three segmentation tasks. These tasks differ in image modalities, complexity, number of structures to be segmented, and spatial and phenotypic heterogeneity. This experimental design was used to underline the potential of D-Net, DLK-Net, and DLK-NETR to generalize across different segmentation tasks. For a thoughtful comparison, we compared their performance with various recent 3D state-of-the-art (SOTA) segmentation models. These models were trained based on the optimal parameters provided in their papers and evaluated by the same 5-fold cross-validation. These methods include
\begin{itemize}
    \item CNN-based methods: VNet (\cite{milletari2016v}), nnU-Net (\cite{isensee2021nnu}), Attention gated U-Net (Att U-Net) (\cite{oktay2018attention})
    \item ViT-based methods: nnFormer (\cite{zhou2023nnformer}), SegFormer (\cite{perera2024segformer3d})
    \item Hybrid CNN-ViT-based methods: TransBTS (\cite{wang2021transbts}), UNETR (\cite{hatamizadeh2022unetr}), Swin UNETR (\cite{hatamizadeh2021swin}), and VSmTrans (\cite{liu2024vsmtrans})
    \item Large convolutional kernel-based methods: 3D UX Net (\cite{lee20223d}) and MedNext (\cite{roy2023mednext})
\end{itemize}

Table~\ref{tab1}, Table~\ref{tab2}, and Table~\ref{tab3} present the performance comparison on the AMOS abdominal multi-organ segmentation task, MSD Brain Tumor segmentation task, and MSD Hepatic Vessel Tumor segmentation task, respectively. Table~\ref{tab4} presents the comparison of architectural complexity. Figure \ref{vis1} shows the qualitative comparison between D-Net and other methods in three tasks. Figure \ref{vis2} shows the qualitative comparison between DLK-NETR and their counterparts in three tasks.

\textbf{AMOS Multi-organ Segmentation.} D-Net achieved an overall average Dice score of $89.67\%$ with the lowest standard deviation, demonstrating the best and most robust overall performance compared with other SOTA methods. Specifically, compared to CNN-based methods, including VNet, nnU-Net, and Att U-Net which have achieved SOTA in various segmentation tasks, D-Net achieved superior overall segmentation performance, while maintaining lower computational complexity, only $39.28$M Params and $200.13$G FLOPs. Additionally, D-Net achieved a significantly higher Dice score than ViT-based methods, including nnFormer and 3D SegFormer. Although SegFormer had a lower computational complexity, D-Net presented a much better segmentation performance. When D-Net was compared with hybrid CNN-ViT methods, it achieved a superior overall performance. Although TransBTS had a lower computational complexity and UNETR had lower FLOPs, D-Net achieved a higher Dice score in the overall segmentation performance. VSmTrans achieved the best performance among these hybrid CNN-ViT methods, but D-Net exceeded it in the overall segmentation performance. 3D UX Net leverages large kernel modules to compete with Transformers. However, D-Net achieved superior performance with much lower computational complexity than it, demonstrating the effectiveness and efficiency of DLK and overall architectures. MedNext had a lower computational complexity, but D-Net demonstrated a higher segmentation accuracy. Additionally, D-Net showed significant improvement in Dice score across all organ-specific segmentation tasks, which demonstrates that D-Net demonstrates superior capabilities to capture features from heterogeneous organs. The superior overall and organ-specific segmentation performance demonstrates the robustness and superior segmentation capabilities of the D-Net. 

DLK-Net achieved an $87.46\%$ overall Dice score, showing superior segmentation performance than other SOTA methods. To be specific, DLK-Net is a lightweight variant of D-Net, and it demonstrated a higher overall segmentation dice score with lower computational complexity than other SOTA methods ($39.12$M Params and $62.37$G FLOPs). Although DLK-Net showed a comparably similar Dice score as nnU-Net and Att U-Net, it had much lower computational complexity than them. SegFormer is a lightweight ViT-based segmentation model and has lower computational complexity, but DLK-Net demonstrated much better performance in overall and organ-specific segmentation results. Thus, DLK-Net can achieve competitive segmentation accuracy efficiently.

DLK-NETR achieved the second-best overall performance in the multi-organ segmentation task ($87.65\%$ with $13.65\%$). DLK-NETR employs the same CNN-ViT architecture as UNETR, Swin UNETR, UX Net, and VSmTrans, so we compared DLK-NETR with them. UNETR and Swin UNETR employ a Transformer encoder and a Swin Transformer encoder, respectively, but DLK-NETR demonstrated a superior and robust segmentation performance with a smaller number of parameters. Thus, employing DLK enhances the capabilities of segmentation networks to capture features from multiple organs with various shapes and sizes, and it demonstrates superior capabilities than Transformer blocks. UX Net employs large convolutional kernels in the encoder, and VSmTrans employs a hybrid module that combines self-attentions and convolutions. However, DLK-NETR employs our DLK to achieve a better segmentation performance with lower computational complexity than them, showing the superior design of our DLK.

\textbf{Brain Tumor Segmentation.} D-Net achieved an average Dice score of $74.42\%$ with a standard deviation of $19.57\%$ in brain tumor segmentation, demonstrating a superior segmentation performance than other SOTA methods. nnU-Net, nnFormer, Swin UNETR, and VSmTrans showed better segmentation performance than other SOTA methods, but D-Net exceeded them in overall and organ-specific performance with lower computational complexity. Although TransBTS and MedNext had fewer Params and FLOPs, D-Net presented higher segmentation accuracy. Additionally, its lightweight variant, DLK-Net, demonstrated much higher segmentation accuracy than another lightweight segmentation method, SegFormer.

DLK-NETR achieved an average Dice score of $74.18\%$, demonstrating superior overall segmentation performance than UNETR, Swin UNETR, UX Net, and VSmTrans. Additionally, although Swin UNETR achieved a higher Dice score in Edema with slightly fewer FLOPs, DLK-NETR achieved higher accuracy in overall segmentation and the segmentation of Enhancing Tumors and Non-Enhancing Tumors with fewer Params. DLK-NETR achieved a much higher segmentation Dice score than UX Net with lower computational complexity, demonstrating the superior segmentation capabilities of the DLK module over the LK module in the UX Net.

\textbf{Hepatic Vessel Tumor Segmentation.} When D-Net was applied to segment hepatic vessels and tumors, it achieved an average $67.63\%$ Dice score, demonstrating superior segmentation performance than other SOTA methods. Additionally, it achieved Dice scores of $65.08\%$ and $70.17\%$ in vessel and tumor segmentation, respectively, which presented superior segmentation capabilities in tubular structures and heterogeneous tumors than other methods. Although MedNext showed lower computational complexity, D-Net demonstrated better overall and organ-specific segmentation performance. Some methods had fewer Params or FLOPs than D-Net, including TransBTS, UNETR, and SegFormer, but DLK-Net was designed as a lightweight variant of D-Net and achieved superior segmentation performance than them ($65.61\%$).

DLK-NETR achieved an average Dice score of $67.75\%$, which is the best overall segmentation performance in hepatic vessels and tumor segmentation. Compared with UNETR, Swin UNETR, UX Net, and VSmTrans, DLK-NETR demonstrated significantly superior overall segmentation performance. Additionally, DLK-NETR achieved much higher Dice scores in both vessel and tumor segmentation tasks than them, demonstrating superior capabilities of the DLK to segment tubular structures and heterogeneous objects efficiently.

\subsection{Ablation Study}
We implemented three ablation studies on the DLK module, DFF module, and the Salience Layer, respectively. In the ablation study, we evaluated the segmentation performance and computational complexity of different architecture configurations on the 2022 AMOS multi-organ segmentation dataset. We first illustrate the details of several backbones that we used in these studies.
\begin{itemize}
    \item nnU-Net: it utilizes a 6-layer U-Net for segmentation. This network consists of a 5-stage encoder, a bottleneck, and a 5-stage decoder. In each stage, two consecutive $3\times3\times3$ convolutional layers are employed for feature extraction.
    \item Conv-ViT: it utilizes the same hierarchical ViT architecture as DLK-Net. However, it employs $5\times5\times5$ depth-wise convolutional layers rather than DLK for generic feature extraction, and it does not employ DFF.
    \item DLK-ViT: it utilizes the same hierarchical ViT architecture as DLK-Net. DLK is utilized for generic feature extraction, but it does not employ DFF.
\end{itemize}

\subsubsection{Ablation on different DLK configurations} 
In this ablation study, we first evaluated the benefits of utilizing two depth-wise convolutional layers to generate a large receptive field. Subsequently, we evaluated the effectiveness of dynamic mechanisms in the DLK. Specifically, in the Conv-ViT backbone, a depth-wise convolutional layer with a $5\times5\times5$ kernel ($5\times5\times5$ DWConv) was employed for feature extraction. Subsequently, $5\times5\times5$ and $7\times7\times7$ depth-wise convolutional layers from DLK were employed to provide a receptive field as a $23\times23\times23$ depth-wise convolutional layer ($23\times23\times23$ DWConv). However, they were stacked and concatenated rather than being fused by dynamic mechanisms. Lastly, DLK was utilized as the basic module and employed in the Conv-ViT backbone. The same configuration was applied to the D-Net backbone to evaluate their benefits. 

\begin{figure}[!t]
\centering
\includegraphics[width=0.48\textwidth]{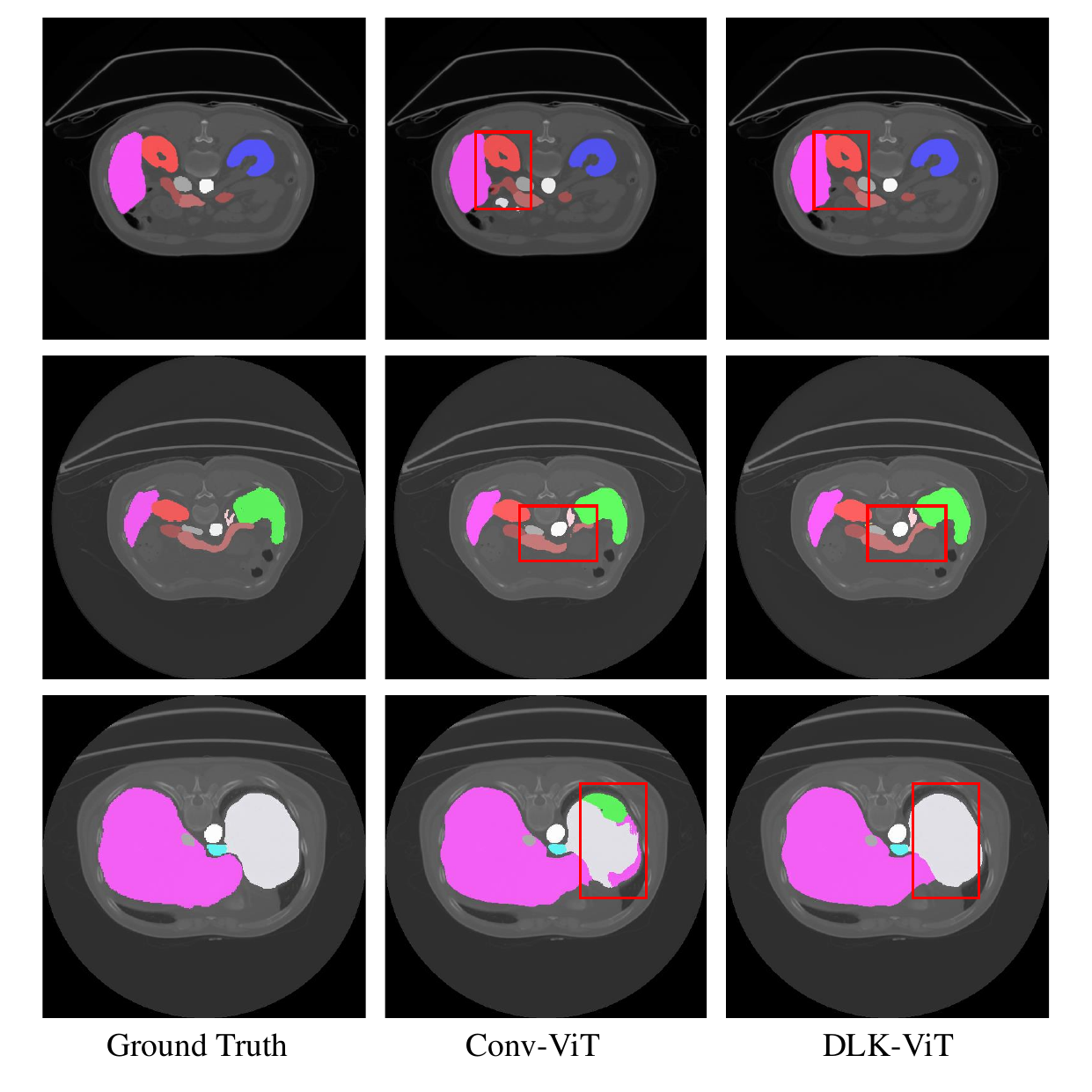}
\caption{Qualitative comparison between Conv-ViT and DLK-ViT on AMOS multi-organ segmentation task. DLK-ViT shows the best segmentation quality than Conv-ViT, demonstrating the effectiveness and superior segmentation capabilities of DLK.} 
\label{vis3}
\end{figure}

Table~\ref{tab5} demonstrates the results of the ablation study on DLK. These results demonstrate the effectiveness of DLK on segmentation. Specifically, when $23\times23\times23$ depth-wise convolutional layers were utilized rather than a $5\times5\times5$ convolutional layer in Conv-ViT and D-Net backbones, the average Dice scores were improved from $84.37\%$ to $85.30\%$ and from $87.62\%$ to $88.20\%$, respectively. These results demonstrate that extracting multi-scale features via a larger receptive field improved segmentation performance. When dynamic mechanisms were employed in DLK, the Dice scores were constantly improved to $86.27\%$ and $89.67\%$ in Conv-ViT and D-Net, respectively. Thus, employing dynamic mechanisms improved the capabilities of DLK to capture global information, enhancing the overall segmentation capabilities of networks. Figure \ref{vis3} demonstrates the qualitative comparison results between Conv-ViT and DLK-ViT. DLK-ViT presents a higher segmentation quality than Conv-ViT, demonstrating the improvement of DLK on segmentation performance. 

DLK is designed as a lightweight module for efficient medical image segmentation, and its incorporation only slightly increased the computational complexity. To be specific, employing DLK in the Conv-ViT increased around $8\%$ Params and FLOPs over employing a $5\times5\times5$ convolutional layer. Similarly, employing DLK in the D-Net increased around $8\%$ in Params and only $2\%$ in FLOPs.

\begin{table}[!t]
\centering
\caption{Ablation study on the effectiveness of DLK on the 2022 AMOS multi-organ segmentation task. The Params and FLOPs are evaluated for input patches with the dimension of $96\times160\times160$.}
\resizebox{0.48\textwidth}{!}{
\begin{tabular}{c|c|cc|cc}
\toprule
Backbones & Basic modules  & Mean Dice & SD Dice & Params & FLOPs \\
\midrule
\multirow{3}{*}{Conv-ViT} & $5\times5\times5$ DWConv    & 84.37 & 16.20 & 34.88 & 159.79 \\
                          & $23\times23\times23$ DWConv & 85.30 & 16.04 & 36.34 & 171.04 \\
                          & DLK                         & 86.27 & 15.52 & 38.33 & 173.13 \\
\midrule
\multirow{3}{*}{D-Net} & $5\times5\times5$ DWConv     & 87.62  & 14.50 & 35.83 & 542.58 \\
                       & $23\times23\times23$ DWConv  & 88.20  & 12.68 & 37.29 & 553.83 \\
                       & DLK                          & 89.67  & 12.56 & 39.28 & 555.91 \\
\bottomrule
\end{tabular}}
\label{tab5}
\end{table}

\subsubsection{Impact of DFF module} 
\begin{figure}[!t]
\centering
\includegraphics[width=0.48\textwidth]{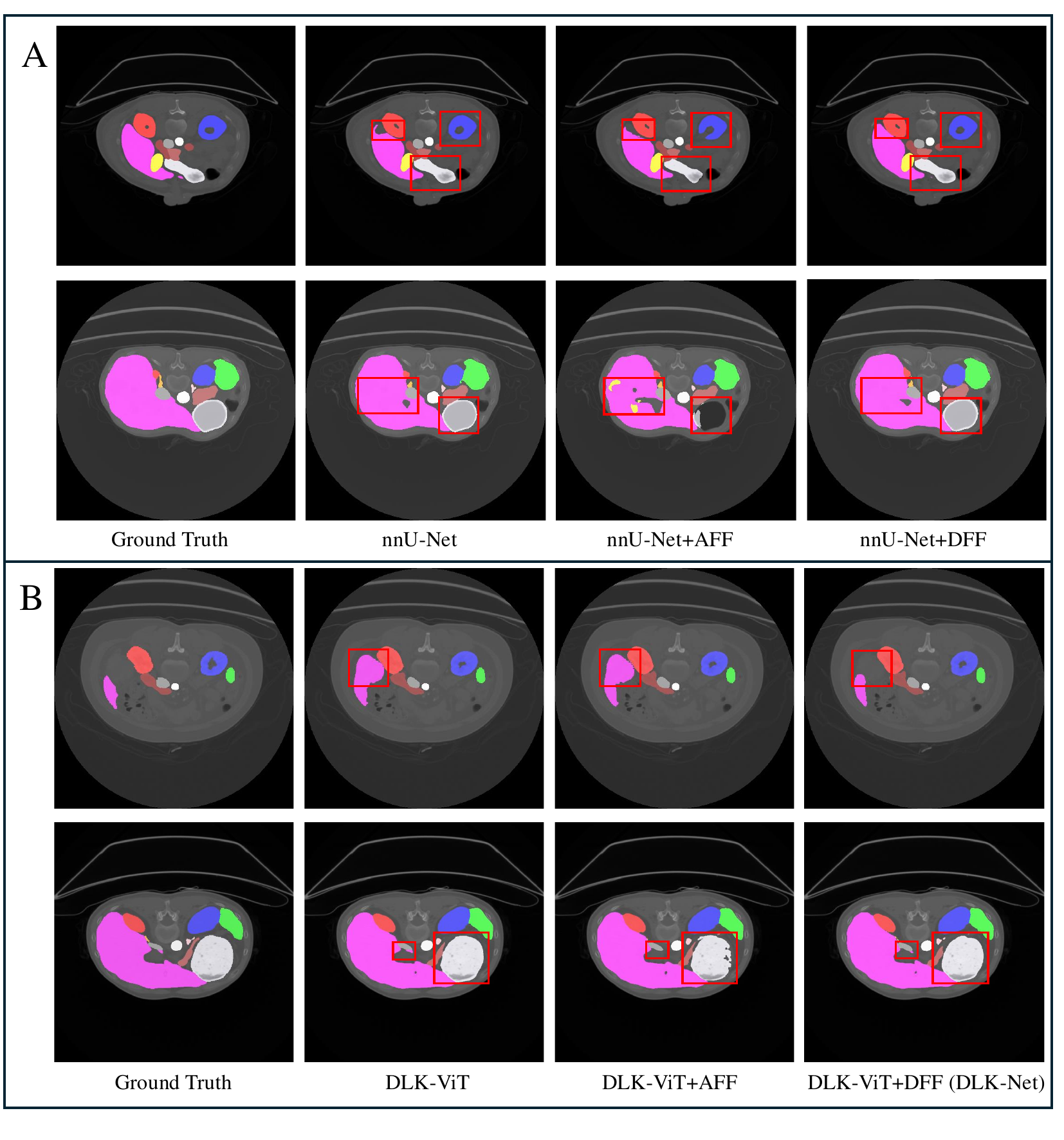}
\caption{(A) Qualitative comparison between nnU-net, nnU-Net incorporated with AFF (nnU-Net+AFF), and nnU-Net incorporated with DFF (nnU-Net+DFF) on AMOS multi-organ segmentation task. (B) Qualitative comparison between DLK-ViT, DLK-ViT incorporated with AFF (DLK-ViT+AFF), and DLK-ViT incorporated with DFF (DLK-ViT+DFF(DLK-Net)) on AMOS multi-organ segmentation task. The incorporation of DFF into nnU-Net and DLK-ViT improves their segmentation quality. Additionally, nnU-Net and DLK-ViT incorporated with DFF show better segmentation quality than those incorporated with AFF.} 
\label{vis4}
\end{figure}

In this study, we investigated the impact and effectiveness of the DFF module. First, we incorporated DFF into three backbones to evaluate its effectiveness in different segmentation architectures, including nnU-Net, Conv-ViT, and DLK-ViT. They are a pure CNN backbone, a hybrid CNN-ViT backbone, and a hybrid LK-ViT backbone respectively. Table \ref{tab6} demonstrates the experimental results of incorporating DFF into these backbones. Figure \ref{vis4} demonstrates the qualitative comparison results of nnU-Net and DLK-ViT with and without DFF. To be specific, the incorporation of DFF into these backbones improved segmentation accuracy by $1-2\%$ Dice scores. Additionally, DFF is designed as a lightweight module, so its incorporation into Conv-ViT and DLK-ViT increased Params and FLOPs slightly, only around $0.8$M and $0.1$G, respectively. Additionally, incorporating DFF into the nnU-Net decreased the Params and FLOPs. It can be explained by the design and functions of the DFF module. DFF is designed to fuse features dynamically, and it will maintain the number of channels unchanged during fusion. Thus, when DFF fuses $C$-channel features from the encoder and $C$-channel upsampled features, it generates output features with the same dimension as upsampled features for following convolutional blocks. However, the original nnU-Net lacks this mechanism, so output features are generated via channel-wise concatenation, and the following convolutional blocks take these $2C$-channel features as inputs, resulting in more Params and FLOPs.

Subsequently, to evaluate the importance of the DFF module on D-Net, we removed DFF from D-Net and utilized the concatenation to fuse upsampled features in the decoder with skip-connected features from the encoder. This fusing strategy is commonly used in various segmentation networks (\cite{ronneberger2015u,isensee2021nnu,hatamizadeh2022unetr,hatamizadeh2021swin}), thus motivating us to design this ablation study. Table \ref{tab6} demonstrates that employing DFF in the D-Net improved the segmentation accuracy from $88.04\%$ to $89.67\%$, while only slightly increasing computational complexity. 

To present the superior performance of the DFF, we compared it with another popular feature fusion module, the Attentional Feature Fusion (AFF) (\cite{dai2021attentional}) module. AFF and DFF were incorporated into three backbones, including nnU-Net, Conv-ViT, and DLK-ViT, to fuse upsampled features in the decoder with skip-connected features from the encoder. Table \ref{tab7} demonstrates the results of these backbones incorporated with AFF and DFF. Figure \ref{vis4} demonstrates the qualitative comparison results of nnU-Net and DLK-ViT incorporated with DFF and AFF. The incorporation of DFF improved the segmentation accuracy of nnU-Net, Conv-ViT, and DLK-ViT by around $2\%$-$4\%$ Dice points compared with AFF. Thus, DFF enhances the segmentation capabilities of these segmentation networks, showing high effectiveness and superior capabilities.

\begin{table}[!t]
\centering
\caption{Ablation study on the impact of DFF module on various backbones on the 2022 AMOS multi-organ segmentation task. The Params and FLOPs are evaluated for input patches with the dimension of $96\times160\times160$.}
\resizebox{0.45\textwidth}{!}{
\begin{tabular}{c|c|cc|cc}
\toprule
Backbones         & DFF & Mean Dice & SD Dice & Params & FLOPs  \\
\midrule
\multirow{2}{*}{nnU-Net}   &            & 87.39 & 14.47 & 68.38 & 992.02  \\
                           & \checkmark & 88.69 & 13.79 & 61.05 & 870.76  \\
\midrule
\multirow{2}{*}{Conv-ViT}  &            & 84.37 & 16.20 & 34.88 & 159.79  \\
                           & \checkmark & 85.62 & 15.29 & 35.66 & 159.89  \\
\midrule
\multirow{2}{*}{DLK-ViT}   &            & 86.27 & 16.04 & 38.33 & 173.13 \\
                           & \checkmark & 87.46 & 14.74 & 39.12 & 173.23  \\
\midrule
\multirow{2}{*}{D-Net}     &            & 88.04 & 12.99 & 38.49 & 555.22  \\
                           & \checkmark & 89.67 & 12.56 & 39.28 & 555.91  \\
\bottomrule
\end{tabular}}
\label{tab6}
\end{table}

\begin{table}[!t]
\centering
\caption{Comparison of segmentation performance and computational complexity among DFF module and another feature fusion module, AFF, on the 2022 AMOS multi-organ segmentation task. The Params and FLOPs are evaluated for input patches with the dimension of $96\times160\times160$.}
\resizebox{0.45\textwidth}{!}{
\begin{tabular}{c|c|cc|cc}
\toprule
Backbones         & Modules & Mean Dice & SD Dice & Params & FLOPs  \\
\midrule
\multirow{2}{*}{nnU-Net}   & AFF & 85.26 & 16.15 & 59.31 & 863.46  \\
                           & DFF & 88.69 & 13.79 & 61.05 & 870.76  \\
\midrule
\multirow{2}{*}{Conv-ViT}  & AFF & 83.82 & 16.59 & 34.69 & 157.91 \\
                           & DFF & 85.62 & 15.29 & 35.66 & 159.89 \\
\midrule
\multirow{2}{*}{DLK-ViT}   & AFF & 85.58 & 14.89 & 38.14 & 171.24 \\
                           & DFF & 87.46 & 14.74 & 39.12 & 173.23 \\
\bottomrule
\end{tabular}}
\label{tab7}
\end{table}

\subsubsection{Comparison of different Salience layers} 
We first investigated the contribution of the Salience layer on segmentation performance by removing it from D-Net. Subsequently, we evaluated the effectiveness of different blocks on the Salience layer, including the convolutional block (ConvBlock), the DLK, and the Channel Mixer. The experimental results in Table \ref{tab8} demonstrate that incorporating the Salience layer into the D-Net improved the segmentation performance of the D-Net by around $1-3\%$ points from $87.46\%$ average Dice score. However, the architectural complexity increased due to its incorporation. Additionally, incorporating the Channel Mixer in the Salience layer achieved higher segmentation accuracy than the ConvBlock ($89.67\%$ vs. $89.34\%$ Dice scores) with lower computational complexity. Additionally, when employing the DLK module in the Salience layer rather than the Channel Mixer, D-Net achieved similar overall segmentation accuracy and computational complexity.

\begin{table}[!t]
\centering
\caption{Ablation study on Salience layer on the 2022 AMOS multi-organ segmentation task. The Params and FLOPs are evaluated for input patches with the dimension of $96\times160\times160$.}
\resizebox{0.45\textwidth}{!}{
\begin{tabular}{c|cc|cc}
\toprule
Saliency Layer  & Mean Dice & SD Dice & Params & FLOPs \\
\midrule
No              & 87.46  & 14.74 & 39.12 & 173.23 \\
ConvBlock       & 89.34  & 12.86 & 39.32 & 648.99 \\
DLK             & 89.59  & 12.58 & 39.28 & 558.24 \\
Channel Mixer   & 89.67  & 12.56 & 39.28 & 555.91  \\
\bottomrule
\end{tabular}}
\label{tab8}
\end{table}

\section{Discussion}
\begin{figure*}[!t]
\centering
\includegraphics[width=\textwidth]{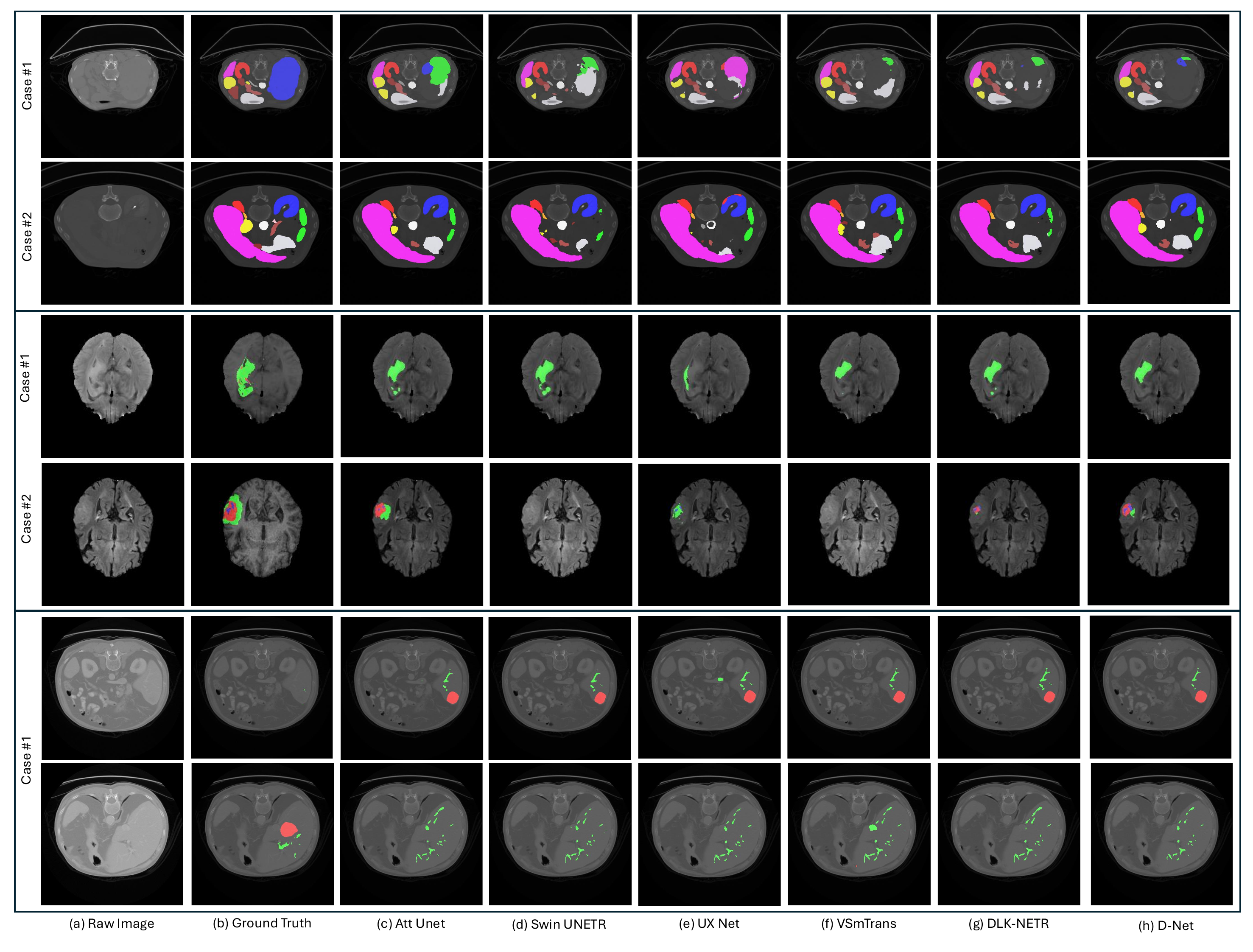}
\caption{Qualitative comparison of failure segmentation results from D-Net, DLK-NETR, and other SOAT methods across three public datasets. Two cases are shown from the AMOS multi-organ dataset and the Brain Tumor dataset, and one case is shown from the Hepatic Vessel Tumor dataset.} 
\label{vis5}
\end{figure*}

To understand the limitations of DLK-NETR and D-Net, we presented a comprehensive analysis of the segmentation results with the lowest Dice scores by these methods along with other SOTA methods across three segmentation tasks. Table \ref{tab9} presents the Dice score of these failure cases, and Figure \ref{vis5} demonstrates these cases. 

Failure cases by these methods exhibited notable consistency, as nearly all methods demonstrated low segmentation accuracy in the same instances across three datasets. Specifically, Att U-Net, Swin UNETR, DLK-NETR, and D-Net demonstrated the lowest segmentation accuracy in the first case and similarly low performance in the second case for the AMOS multi-organ segmentation. In contrast, UX Net and VSmTrans demonstrated the lowest segmentation Dice score in the second case while underperforming in the first case. In the Brain Tumor segmentation task, Att U-Net, DLK-NETR, and D-Net demonstrated the lowest segmentation accuracy in the first case and showed low segmentation performance in the second case. In contrast, Swin UNETR and VSmTrans demonstrated the lowest segmentation Dice score in the second case while underperforming in the first case. In this study, we have included various architectures, including CNN (Att U-Net), hybrid CNN-ViT (Swin UNETR and VSmTrans), LK (UX Net and DLK-NETR), and hybrid LK-ViT architectures (D-Net). Thus, these limitations may not be from the architectural designs or training protocols. We hypothesize that certain anatomical features or pathological characteristics may inherently pose difficulties for segmentation algorithms.

Additionally, all segmentation methods exhibited the lowest segmentation accuracy in the same instance of the MSD Vessel and Tumor segmentation dataset, achieving approximately a $1\%$ Dice score. After a thorough examination of the segmentation results and the ground truth mask, we hypothesize that there is a misalignment between the raw image and its corresponding annotation for this specific case (MSD Vessel Tumor; Case ID: $hepaticvessel\_075$).

\begin{table}[!t]
\centering
\caption{Comparison of failure segmentation results from D-Net, DLK-NETR, and other methods on three segmentation tasks. These segmentation results are evaluated by the average Dice score on each case. ($^*$ indicates the case that has the lowest Dice score on this dataset by each method.)}
\resizebox{0.45\textwidth}{!}{
\begin{tabular}{c|cc|cc|c}
\toprule
  & \multicolumn{2}{c|}{AMOS Multi-organ} & \multicolumn{2}{c|}{Brain Tumor} & Vessel Tumor \\
\midrule
Methods & Case $\#1$ & Case $\#2$   & Case $\#1$ & Case $\#2$ & Case $\#1$ \\
\midrule
Att U-Net   & 61.37$^*$ & 65.19     & 21.46$^*$ & 33.94     & 0.86$^*$ \\
Swin UNETR & 59.37$^*$ & 60.10     & 22.26     & 0.00$^*$  & 1.02$^*$ \\
UX Net     & 55.40     & 52.37$^*$ & 12.74     & 16.40     & 0.99$^*$ \\
VSmTrans   & 60.24     & 59.33$^*$ & 14.81     & 0.16$^*$  & 0.78$^*$ \\
\midrule
DLK-NETR  & 63.10$^*$  & 63.96     & 19.58$^*$ & 20.81     & 0.85$^*$ \\
D-Net     & 67.12$^*$  & 68.91     & 16.34$^*$ & 34.12     & 0.90$^*$ \\
\bottomrule
\end{tabular}}
\label{tab9}
\end{table}

\section{Conclusion}
We introduced D-Net for volumetric medical image segmentation by incorporating a Dynamic Large Kernel and Dynamic Feature Fusion modules, along with a Salience layer, into a hierarchical transformer architecture. DLK is proposed to employ multiple large convolutional kernels to capture richer contextual information via a large receptive field and utilize dynamic mechanisms to adjust these features adaptively based on global information. DFF is proposed to adaptively fuse features from different levels. The Salience layer is designed to capture more low-level features to enhance the segmentation capabilities of D-Net. Two other variants, DLK-Net and DLK-NETR were proposed for efficient medical image segmentation. We evaluated D-Net, DLK-Net, and DLK-NETR on three heterogeneous segmentation tasks, and they demonstrated superior segmentation performance with lower computation complexity compared to SOTA methods. Due to their high generalizability and applicability, we believe that D-Net has the potential to achieve promising segmentation performance on various medical image segmentation tasks.

\section*{Declaration of competing interests}
The authors declare that they have no known competing financial interests or personal relationships that could have appeared to influence the work reported in this paper.

\section*{Acknowledgments}
This work was supported by NIH grant U24 CA258483. Computations were performed using the facilities of the Washington University Research Computing and Informatics Facility (RCIF). The RCIF has received funding from NIH S10 program grants: 1S10OD025200-01A1 and 1S10OD030477-01.

\section*{CRediT authorship contribution statement}
JY: conceptualization, methodology, formal analysis, writing the original draft, reviewing, and editing, visualization; PQ: conceptualization; YZ: reviewing; DM: methodology, writing, reviewing, and editing, supervision; AS: methodology, writing, reviewing, and editing, supervision.








\bibliographystyle{cas-model2-names}

\bibliography{DNet}

\begin{thebibliography}{54}
\expandafter\ifx\csname natexlab\endcsname\relax\def\natexlab#1{#1}\fi
\providecommand{\url}[1]{\texttt{#1}}
\providecommand{\href}[2]{#2}
\providecommand{\path}[1]{#1}
\providecommand{\DOIprefix}{doi:}
\providecommand{\ArXivprefix}{arXiv:}
\providecommand{\URLprefix}{URL: }
\providecommand{\Pubmedprefix}{pmid:}
\providecommand{\doi}[1]{\href{http://dx.doi.org/#1}{\path{#1}}}
\providecommand{\Pubmed}[1]{\href{pmid:#1}{\path{#1}}}
\providecommand{\bibinfo}[2]{#2}
\ifx\xfnm\relax \def\xfnm[#1]{\unskip,\space#1}\fi
\bibitem[{Antonelli et~al.(2022)Antonelli, Reinke, Bakas, Farahani, Kopp-Schneider, Landman, Litjens, Menze, Ronneberger, Summers et~al.}]{antonelli2022medical}
\bibinfo{author}{Antonelli, M.}, \bibinfo{author}{Reinke, A.}, \bibinfo{author}{Bakas, S.}, \bibinfo{author}{Farahani, K.}, \bibinfo{author}{Kopp-Schneider, A.}, \bibinfo{author}{Landman, B.A.}, \bibinfo{author}{Litjens, G.}, \bibinfo{author}{Menze, B.}, \bibinfo{author}{Ronneberger, O.}, \bibinfo{author}{Summers, R.M.}, et~al., \bibinfo{year}{2022}.
\newblock \bibinfo{title}{The medical segmentation decathlon}.
\newblock \bibinfo{journal}{Nature communications} \bibinfo{volume}{13}, \bibinfo{pages}{4128}.
\bibitem[{Azad et~al.(2024)Azad, Niggemeier, H{\"u}ttemann, Kazerouni, Aghdam, Velichko, Bagci and Merhof}]{azad2024beyond}
\bibinfo{author}{Azad, R.}, \bibinfo{author}{Niggemeier, L.}, \bibinfo{author}{H{\"u}ttemann, M.}, \bibinfo{author}{Kazerouni, A.}, \bibinfo{author}{Aghdam, E.K.}, \bibinfo{author}{Velichko, Y.}, \bibinfo{author}{Bagci, U.}, \bibinfo{author}{Merhof, D.}, \bibinfo{year}{2024}.
\newblock \bibinfo{title}{Beyond self-attention: Deformable large kernel attention for medical image segmentation}, in: \bibinfo{booktitle}{Proceedings of the IEEE/CVF Winter Conference on Applications of Computer Vision}, pp. \bibinfo{pages}{1287--1297}.
\bibitem[{Cao et~al.(2022)Cao, Wang, Chen, Jiang, Zhang, Tian and Wang}]{cao2022swin}
\bibinfo{author}{Cao, H.}, \bibinfo{author}{Wang, Y.}, \bibinfo{author}{Chen, J.}, \bibinfo{author}{Jiang, D.}, \bibinfo{author}{Zhang, X.}, \bibinfo{author}{Tian, Q.}, \bibinfo{author}{Wang, M.}, \bibinfo{year}{2022}.
\newblock \bibinfo{title}{Swin-unet: Unet-like pure transformer for medical image segmentation}, in: \bibinfo{booktitle}{European conference on computer vision}, \bibinfo{organization}{Springer}. pp. \bibinfo{pages}{205--218}.
\bibitem[{Chen et~al.(2021)Chen, Lu, Yu, Luo, Adeli, Wang, Lu, Yuille and Zhou}]{chen2021transunet}
\bibinfo{author}{Chen, J.}, \bibinfo{author}{Lu, Y.}, \bibinfo{author}{Yu, Q.}, \bibinfo{author}{Luo, X.}, \bibinfo{author}{Adeli, E.}, \bibinfo{author}{Wang, Y.}, \bibinfo{author}{Lu, L.}, \bibinfo{author}{Yuille, A.L.}, \bibinfo{author}{Zhou, Y.}, \bibinfo{year}{2021}.
\newblock \bibinfo{title}{Transunet: Transformers make strong encoders for medical image segmentation}.
\newblock \bibinfo{journal}{arXiv preprint arXiv:2102.04306} .
\bibitem[{Chen et~al.(2023)Chen, Mei, Li, Lu, Yu, Wei, Luo, Xie, Adeli, Wang et~al.}]{chen20233d}
\bibinfo{author}{Chen, J.}, \bibinfo{author}{Mei, J.}, \bibinfo{author}{Li, X.}, \bibinfo{author}{Lu, Y.}, \bibinfo{author}{Yu, Q.}, \bibinfo{author}{Wei, Q.}, \bibinfo{author}{Luo, X.}, \bibinfo{author}{Xie, Y.}, \bibinfo{author}{Adeli, E.}, \bibinfo{author}{Wang, Y.}, et~al., \bibinfo{year}{2023}.
\newblock \bibinfo{title}{3d transunet: Advancing medical image segmentation through vision transformers}.
\newblock \bibinfo{journal}{arXiv preprint arXiv:2310.07781} .
\bibitem[{Chen et~al.(2017)Chen, Papandreou, Kokkinos, Murphy and Yuille}]{chen2017deeplab}
\bibinfo{author}{Chen, L.C.}, \bibinfo{author}{Papandreou, G.}, \bibinfo{author}{Kokkinos, I.}, \bibinfo{author}{Murphy, K.}, \bibinfo{author}{Yuille, A.L.}, \bibinfo{year}{2017}.
\newblock \bibinfo{title}{Deeplab: Semantic image segmentation with deep convolutional nets, atrous convolution, and fully connected crfs}.
\newblock \bibinfo{journal}{IEEE transactions on pattern analysis and machine intelligence} \bibinfo{volume}{40}, \bibinfo{pages}{834--848}.
\bibitem[{Chen et~al.(2022)Chen, Wang, Zhang, Fung, Thai, Moore, Mannel, Liu, Zheng and Qiu}]{chen2022recent}
\bibinfo{author}{Chen, X.}, \bibinfo{author}{Wang, X.}, \bibinfo{author}{Zhang, K.}, \bibinfo{author}{Fung, K.M.}, \bibinfo{author}{Thai, T.C.}, \bibinfo{author}{Moore, K.}, \bibinfo{author}{Mannel, R.S.}, \bibinfo{author}{Liu, H.}, \bibinfo{author}{Zheng, B.}, \bibinfo{author}{Qiu, Y.}, \bibinfo{year}{2022}.
\newblock \bibinfo{title}{Recent advances and clinical applications of deep learning in medical image analysis}.
\newblock \bibinfo{journal}{Medical image analysis} \bibinfo{volume}{79}, \bibinfo{pages}{102444}.
\bibitem[{Chen et~al.(2020)Chen, Dai, Liu, Chen, Yuan and Liu}]{chen2020dynamic}
\bibinfo{author}{Chen, Y.}, \bibinfo{author}{Dai, X.}, \bibinfo{author}{Liu, M.}, \bibinfo{author}{Chen, D.}, \bibinfo{author}{Yuan, L.}, \bibinfo{author}{Liu, Z.}, \bibinfo{year}{2020}.
\newblock \bibinfo{title}{Dynamic convolution: Attention over convolution kernels}, in: \bibinfo{booktitle}{Proceedings of the IEEE/CVF conference on computer vision and pattern recognition}, pp. \bibinfo{pages}{11030--11039}.
\bibitem[{Dai et~al.(2021)Dai, Gieseke, Oehmcke, Wu and Barnard}]{dai2021attentional}
\bibinfo{author}{Dai, Y.}, \bibinfo{author}{Gieseke, F.}, \bibinfo{author}{Oehmcke, S.}, \bibinfo{author}{Wu, Y.}, \bibinfo{author}{Barnard, K.}, \bibinfo{year}{2021}.
\newblock \bibinfo{title}{Attentional feature fusion}, in: \bibinfo{booktitle}{Proceedings of the IEEE/CVF winter conference on applications of computer vision}, pp. \bibinfo{pages}{3560--3569}.
\bibitem[{Ding et~al.(2022)Ding, Zhang, Han and Ding}]{ding2022scaling}
\bibinfo{author}{Ding, X.}, \bibinfo{author}{Zhang, X.}, \bibinfo{author}{Han, J.}, \bibinfo{author}{Ding, G.}, \bibinfo{year}{2022}.
\newblock \bibinfo{title}{Scaling up your kernels to 31x31: Revisiting large kernel design in cnns}, in: \bibinfo{booktitle}{Proceedings of the IEEE/CVF conference on computer vision and pattern recognition}, pp. \bibinfo{pages}{11963--11975}.
\bibitem[{Dosovitskiy et~al.(2020)Dosovitskiy, Beyer, Kolesnikov, Weissenborn, Zhai, Unterthiner, Dehghani, Minderer, Heigold, Gelly et~al.}]{dosovitskiy2020image}
\bibinfo{author}{Dosovitskiy, A.}, \bibinfo{author}{Beyer, L.}, \bibinfo{author}{Kolesnikov, A.}, \bibinfo{author}{Weissenborn, D.}, \bibinfo{author}{Zhai, X.}, \bibinfo{author}{Unterthiner, T.}, \bibinfo{author}{Dehghani, M.}, \bibinfo{author}{Minderer, M.}, \bibinfo{author}{Heigold, G.}, \bibinfo{author}{Gelly, S.}, et~al., \bibinfo{year}{2020}.
\newblock \bibinfo{title}{An image is worth 16x16 words: Transformers for image recognition at scale}.
\newblock \bibinfo{journal}{arXiv preprint arXiv:2010.11929} .
\bibitem[{Graham et~al.(2021)Graham, El-Nouby, Touvron, Stock, Joulin, J{\'e}gou and Douze}]{graham2021levit}
\bibinfo{author}{Graham, B.}, \bibinfo{author}{El-Nouby, A.}, \bibinfo{author}{Touvron, H.}, \bibinfo{author}{Stock, P.}, \bibinfo{author}{Joulin, A.}, \bibinfo{author}{J{\'e}gou, H.}, \bibinfo{author}{Douze, M.}, \bibinfo{year}{2021}.
\newblock \bibinfo{title}{Levit: a vision transformer in convnet's clothing for faster inference}, in: \bibinfo{booktitle}{Proceedings of the IEEE/CVF international conference on computer vision}, pp. \bibinfo{pages}{12259--12269}.
\bibitem[{Guo et~al.(2022)Guo, Lu, Hou, Liu, Cheng and Hu}]{guo2022segnext}
\bibinfo{author}{Guo, M.H.}, \bibinfo{author}{Lu, C.Z.}, \bibinfo{author}{Hou, Q.}, \bibinfo{author}{Liu, Z.}, \bibinfo{author}{Cheng, M.M.}, \bibinfo{author}{Hu, S.M.}, \bibinfo{year}{2022}.
\newblock \bibinfo{title}{Segnext: Rethinking convolutional attention design for semantic segmentation}.
\newblock \bibinfo{journal}{Advances in Neural Information Processing Systems} \bibinfo{volume}{35}, \bibinfo{pages}{1140--1156}.
\bibitem[{Han et~al.(2021)Han, Huang, Song, Yang, Wang and Wang}]{han2021dynamic}
\bibinfo{author}{Han, Y.}, \bibinfo{author}{Huang, G.}, \bibinfo{author}{Song, S.}, \bibinfo{author}{Yang, L.}, \bibinfo{author}{Wang, H.}, \bibinfo{author}{Wang, Y.}, \bibinfo{year}{2021}.
\newblock \bibinfo{title}{Dynamic neural networks: A survey}.
\newblock \bibinfo{journal}{IEEE Transactions on Pattern Analysis and Machine Intelligence} \bibinfo{volume}{44}, \bibinfo{pages}{7436--7456}.
\bibitem[{Hatamizadeh et~al.(2021)Hatamizadeh, Nath, Tang, Yang, Roth and Xu}]{hatamizadeh2021swin}
\bibinfo{author}{Hatamizadeh, A.}, \bibinfo{author}{Nath, V.}, \bibinfo{author}{Tang, Y.}, \bibinfo{author}{Yang, D.}, \bibinfo{author}{Roth, H.R.}, \bibinfo{author}{Xu, D.}, \bibinfo{year}{2021}.
\newblock \bibinfo{title}{Swin unetr: Swin transformers for semantic segmentation of brain tumors in mri images}, in: \bibinfo{booktitle}{International MICCAI Brainlesion Workshop}, \bibinfo{organization}{Springer}. pp. \bibinfo{pages}{272--284}.
\bibitem[{Hatamizadeh et~al.(2022a)Hatamizadeh, Tang, Nath, Yang, Myronenko, Landman, Roth and Xu}]{hatamizadeh2022unetr}
\bibinfo{author}{Hatamizadeh, A.}, \bibinfo{author}{Tang, Y.}, \bibinfo{author}{Nath, V.}, \bibinfo{author}{Yang, D.}, \bibinfo{author}{Myronenko, A.}, \bibinfo{author}{Landman, B.}, \bibinfo{author}{Roth, H.R.}, \bibinfo{author}{Xu, D.}, \bibinfo{year}{2022}a.
\newblock \bibinfo{title}{Unetr: Transformers for 3d medical image segmentation}, in: \bibinfo{booktitle}{Proceedings of the IEEE/CVF winter conference on applications of computer vision}, pp. \bibinfo{pages}{574--584}.
\bibitem[{Hatamizadeh et~al.(2022b)Hatamizadeh, Xu, Yang, Li, Roth and Xu}]{hatamizadeh2022unetformer}
\bibinfo{author}{Hatamizadeh, A.}, \bibinfo{author}{Xu, Z.}, \bibinfo{author}{Yang, D.}, \bibinfo{author}{Li, W.}, \bibinfo{author}{Roth, H.}, \bibinfo{author}{Xu, D.}, \bibinfo{year}{2022}b.
\newblock \bibinfo{title}{Unetformer: A unified vision transformer model and pre-training framework for 3d medical image segmentation}.
\newblock \bibinfo{journal}{arXiv preprint arXiv:2204.00631} .
\bibitem[{Heidari et~al.(2023)Heidari, Kazerouni, Soltany, Azad, Aghdam, Cohen-Adad and Merhof}]{heidari2023hiformer}
\bibinfo{author}{Heidari, M.}, \bibinfo{author}{Kazerouni, A.}, \bibinfo{author}{Soltany, M.}, \bibinfo{author}{Azad, R.}, \bibinfo{author}{Aghdam, E.K.}, \bibinfo{author}{Cohen-Adad, J.}, \bibinfo{author}{Merhof, D.}, \bibinfo{year}{2023}.
\newblock \bibinfo{title}{Hiformer: Hierarchical multi-scale representations using transformers for medical image segmentation}, in: \bibinfo{booktitle}{Proceedings of the IEEE/CVF winter conference on applications of computer vision}, pp. \bibinfo{pages}{6202--6212}.
\bibitem[{Hu et~al.(2018)Hu, Shen and Sun}]{hu2018squeeze}
\bibinfo{author}{Hu, J.}, \bibinfo{author}{Shen, L.}, \bibinfo{author}{Sun, G.}, \bibinfo{year}{2018}.
\newblock \bibinfo{title}{Squeeze-and-excitation networks}, in: \bibinfo{booktitle}{Proceedings of the IEEE conference on computer vision and pattern recognition}, pp. \bibinfo{pages}{7132--7141}.
\bibitem[{Huang et~al.(2021)Huang, Deng, Li and Yuan}]{huang2021missformer}
\bibinfo{author}{Huang, X.}, \bibinfo{author}{Deng, Z.}, \bibinfo{author}{Li, D.}, \bibinfo{author}{Yuan, X.}, \bibinfo{year}{2021}.
\newblock \bibinfo{title}{Missformer: An effective medical image segmentation transformer}.
\newblock \bibinfo{journal}{arXiv preprint arXiv:2109.07162} .
\bibitem[{Huang et~al.(2022)Huang, Deng, Li, Yuan and Fu}]{huang2022missformer}
\bibinfo{author}{Huang, X.}, \bibinfo{author}{Deng, Z.}, \bibinfo{author}{Li, D.}, \bibinfo{author}{Yuan, X.}, \bibinfo{author}{Fu, Y.}, \bibinfo{year}{2022}.
\newblock \bibinfo{title}{Missformer: An effective transformer for 2d medical image segmentation}.
\newblock \bibinfo{journal}{IEEE Transactions on Medical Imaging} \bibinfo{volume}{42}, \bibinfo{pages}{1484--1494}.
\bibitem[{Isensee et~al.(2021)Isensee, Jaeger, Kohl, Petersen and Maier-Hein}]{isensee2021nnu}
\bibinfo{author}{Isensee, F.}, \bibinfo{author}{Jaeger, P.F.}, \bibinfo{author}{Kohl, S.A.}, \bibinfo{author}{Petersen, J.}, \bibinfo{author}{Maier-Hein, K.H.}, \bibinfo{year}{2021}.
\newblock \bibinfo{title}{nnu-net: a self-configuring method for deep learning-based biomedical image segmentation}.
\newblock \bibinfo{journal}{Nature methods} \bibinfo{volume}{18}, \bibinfo{pages}{203--211}.
\bibitem[{Ji et~al.(2022)Ji, Bai, Ge, Yang, Zhu, Zhang, Li, Zhanng, Ma, Wan et~al.}]{ji2022amos}
\bibinfo{author}{Ji, Y.}, \bibinfo{author}{Bai, H.}, \bibinfo{author}{Ge, C.}, \bibinfo{author}{Yang, J.}, \bibinfo{author}{Zhu, Y.}, \bibinfo{author}{Zhang, R.}, \bibinfo{author}{Li, Z.}, \bibinfo{author}{Zhanng, L.}, \bibinfo{author}{Ma, W.}, \bibinfo{author}{Wan, X.}, et~al., \bibinfo{year}{2022}.
\newblock \bibinfo{title}{Amos: A large-scale abdominal multi-organ benchmark for versatile medical image segmentation}.
\newblock \bibinfo{journal}{Advances in Neural Information Processing Systems} \bibinfo{volume}{35}, \bibinfo{pages}{36722--36732}.
\bibitem[{Lee et~al.(2022)Lee, Bao, Huo and Landman}]{lee20223d}
\bibinfo{author}{Lee, H.H.}, \bibinfo{author}{Bao, S.}, \bibinfo{author}{Huo, Y.}, \bibinfo{author}{Landman, B.A.}, \bibinfo{year}{2022}.
\newblock \bibinfo{title}{3d ux-net: A large kernel volumetric convnet modernizing hierarchical transformer for medical image segmentation}.
\newblock \bibinfo{journal}{arXiv preprint arXiv:2209.15076} .
\bibitem[{Li et~al.(2019)Li, Wang, Hu and Yang}]{li2019selective}
\bibinfo{author}{Li, X.}, \bibinfo{author}{Wang, W.}, \bibinfo{author}{Hu, X.}, \bibinfo{author}{Yang, J.}, \bibinfo{year}{2019}.
\newblock \bibinfo{title}{Selective kernel networks}, in: \bibinfo{booktitle}{Proceedings of the IEEE/CVF conference on computer vision and pattern recognition}, pp. \bibinfo{pages}{510--519}.
\bibitem[{Li et~al.(2023)Li, Hou, Zheng, Cheng, Yang and Li}]{li2023large}
\bibinfo{author}{Li, Y.}, \bibinfo{author}{Hou, Q.}, \bibinfo{author}{Zheng, Z.}, \bibinfo{author}{Cheng, M.M.}, \bibinfo{author}{Yang, J.}, \bibinfo{author}{Li, X.}, \bibinfo{year}{2023}.
\newblock \bibinfo{title}{Large selective kernel network for remote sensing object detection}.
\newblock \bibinfo{journal}{arXiv preprint arXiv:2303.09030} .
\bibitem[{Lin et~al.(2022)Lin, Chen, Xu, Zhang, Lu and Zhang}]{lin2022ds}
\bibinfo{author}{Lin, A.}, \bibinfo{author}{Chen, B.}, \bibinfo{author}{Xu, J.}, \bibinfo{author}{Zhang, Z.}, \bibinfo{author}{Lu, G.}, \bibinfo{author}{Zhang, D.}, \bibinfo{year}{2022}.
\newblock \bibinfo{title}{Ds-transunet: Dual swin transformer u-net for medical image segmentation}.
\newblock \bibinfo{journal}{IEEE Transactions on Instrumentation and Measurement} \bibinfo{volume}{71}, \bibinfo{pages}{1--15}.
\bibitem[{Liu et~al.(2022a)Liu, Chen, Chen, Chen, Xiao, Wu, K{\"a}rkk{\"a}inen, Pechenizkiy, Mocanu and Wang}]{liu2022more}
\bibinfo{author}{Liu, S.}, \bibinfo{author}{Chen, T.}, \bibinfo{author}{Chen, X.}, \bibinfo{author}{Chen, X.}, \bibinfo{author}{Xiao, Q.}, \bibinfo{author}{Wu, B.}, \bibinfo{author}{K{\"a}rkk{\"a}inen, T.}, \bibinfo{author}{Pechenizkiy, M.}, \bibinfo{author}{Mocanu, D.}, \bibinfo{author}{Wang, Z.}, \bibinfo{year}{2022}a.
\newblock \bibinfo{title}{More convnets in the 2020s: Scaling up kernels beyond 51x51 using sparsity}.
\newblock \bibinfo{journal}{arXiv preprint arXiv:2207.03620} .
\bibitem[{Liu et~al.(2024)Liu, Bai, Torigian, Tong and Udupa}]{liu2024vsmtrans}
\bibinfo{author}{Liu, T.}, \bibinfo{author}{Bai, Q.}, \bibinfo{author}{Torigian, D.A.}, \bibinfo{author}{Tong, Y.}, \bibinfo{author}{Udupa, J.K.}, \bibinfo{year}{2024}.
\newblock \bibinfo{title}{Vsmtrans: A hybrid paradigm integrating self-attention and convolution for 3d medical image segmentation}.
\newblock \bibinfo{journal}{Medical Image Analysis} , \bibinfo{pages}{103295}.
\bibitem[{Liu et~al.(2021)Liu, Lin, Cao, Hu, Wei, Zhang, Lin and Guo}]{liu2021swin}
\bibinfo{author}{Liu, Z.}, \bibinfo{author}{Lin, Y.}, \bibinfo{author}{Cao, Y.}, \bibinfo{author}{Hu, H.}, \bibinfo{author}{Wei, Y.}, \bibinfo{author}{Zhang, Z.}, \bibinfo{author}{Lin, S.}, \bibinfo{author}{Guo, B.}, \bibinfo{year}{2021}.
\newblock \bibinfo{title}{Swin transformer: Hierarchical vision transformer using shifted windows}, in: \bibinfo{booktitle}{Proceedings of the IEEE/CVF international conference on computer vision}, pp. \bibinfo{pages}{10012--10022}.
\bibitem[{Liu et~al.(2022b)Liu, Mao, Wu, Feichtenhofer, Darrell and Xie}]{liu2022convnet}
\bibinfo{author}{Liu, Z.}, \bibinfo{author}{Mao, H.}, \bibinfo{author}{Wu, C.Y.}, \bibinfo{author}{Feichtenhofer, C.}, \bibinfo{author}{Darrell, T.}, \bibinfo{author}{Xie, S.}, \bibinfo{year}{2022}b.
\newblock \bibinfo{title}{A convnet for the 2020s}, in: \bibinfo{booktitle}{Proceedings of the IEEE/CVF conference on computer vision and pattern recognition}, pp. \bibinfo{pages}{11976--11986}.
\bibitem[{Luo et~al.(2016)Luo, Li, Urtasun and Zemel}]{luo2016understanding}
\bibinfo{author}{Luo, W.}, \bibinfo{author}{Li, Y.}, \bibinfo{author}{Urtasun, R.}, \bibinfo{author}{Zemel, R.}, \bibinfo{year}{2016}.
\newblock \bibinfo{title}{Understanding the effective receptive field in deep convolutional neural networks}.
\newblock \bibinfo{journal}{Advances in neural information processing systems} \bibinfo{volume}{29}.
\bibitem[{Milletari et~al.(2016)Milletari, Navab and Ahmadi}]{milletari2016v}
\bibinfo{author}{Milletari, F.}, \bibinfo{author}{Navab, N.}, \bibinfo{author}{Ahmadi, S.A.}, \bibinfo{year}{2016}.
\newblock \bibinfo{title}{V-net: Fully convolutional neural networks for volumetric medical image segmentation}, in: \bibinfo{booktitle}{2016 fourth international conference on 3D vision (3DV)}, \bibinfo{organization}{Ieee}. pp. \bibinfo{pages}{565--571}.
\bibitem[{Oktay et~al.(2018)Oktay, Schlemper, Folgoc, Lee, Heinrich, Misawa, Mori, McDonagh, Hammerla, Kainz et~al.}]{oktay2018attention}
\bibinfo{author}{Oktay, O.}, \bibinfo{author}{Schlemper, J.}, \bibinfo{author}{Folgoc, L.L.}, \bibinfo{author}{Lee, M.}, \bibinfo{author}{Heinrich, M.}, \bibinfo{author}{Misawa, K.}, \bibinfo{author}{Mori, K.}, \bibinfo{author}{McDonagh, S.}, \bibinfo{author}{Hammerla, N.Y.}, \bibinfo{author}{Kainz, B.}, et~al., \bibinfo{year}{2018}.
\newblock \bibinfo{title}{Attention u-net: Learning where to look for the pancreas}.
\newblock \bibinfo{journal}{arXiv preprint arXiv:1804.03999} .
\bibitem[{Peiris et~al.(2022)Peiris, Hayat, Chen, Egan and Harandi}]{peiris2022robust}
\bibinfo{author}{Peiris, H.}, \bibinfo{author}{Hayat, M.}, \bibinfo{author}{Chen, Z.}, \bibinfo{author}{Egan, G.}, \bibinfo{author}{Harandi, M.}, \bibinfo{year}{2022}.
\newblock \bibinfo{title}{A robust volumetric transformer for accurate 3d tumor segmentation}, in: \bibinfo{booktitle}{International conference on medical image computing and computer-assisted intervention}, \bibinfo{organization}{Springer}. pp. \bibinfo{pages}{162--172}.
\bibitem[{Perera et~al.(2024)Perera, Navard and Yilmaz}]{perera2024segformer3d}
\bibinfo{author}{Perera, S.}, \bibinfo{author}{Navard, P.}, \bibinfo{author}{Yilmaz, A.}, \bibinfo{year}{2024}.
\newblock \bibinfo{title}{Segformer3d: an efficient transformer for 3d medical image segmentation}, in: \bibinfo{booktitle}{Proceedings of the IEEE/CVF Conference on Computer Vision and Pattern Recognition}, pp. \bibinfo{pages}{4981--4988}.
\bibitem[{Qiu et~al.(2024)Qiu, Yang, Kumar, Ghosh and Sotiras}]{qiu2024agileformer}
\bibinfo{author}{Qiu, P.}, \bibinfo{author}{Yang, J.}, \bibinfo{author}{Kumar, S.}, \bibinfo{author}{Ghosh, S.S.}, \bibinfo{author}{Sotiras, A.}, \bibinfo{year}{2024}.
\newblock \bibinfo{title}{Agileformer: Spatially agile transformer unet for medical image segmentation}.
\newblock \bibinfo{journal}{arXiv preprint arXiv:2404.00122} .
\bibitem[{Ren et~al.(2022)Ren, Li, Wang, Xiao, Du, Liang and Chang}]{ren2022beyond}
\bibinfo{author}{Ren, P.}, \bibinfo{author}{Li, C.}, \bibinfo{author}{Wang, G.}, \bibinfo{author}{Xiao, Y.}, \bibinfo{author}{Du, Q.}, \bibinfo{author}{Liang, X.}, \bibinfo{author}{Chang, X.}, \bibinfo{year}{2022}.
\newblock \bibinfo{title}{Beyond fixation: Dynamic window visual transformer}, in: \bibinfo{booktitle}{Proceedings of the IEEE/CVF Conference on Computer Vision and Pattern Recognition}, pp. \bibinfo{pages}{11987--11997}.
\bibitem[{Ronneberger et~al.(2015)Ronneberger, Fischer and Brox}]{ronneberger2015u}
\bibinfo{author}{Ronneberger, O.}, \bibinfo{author}{Fischer, P.}, \bibinfo{author}{Brox, T.}, \bibinfo{year}{2015}.
\newblock \bibinfo{title}{U-net: Convolutional networks for biomedical image segmentation}, in: \bibinfo{booktitle}{Medical Image Computing and Computer-Assisted Intervention--MICCAI 2015: 18th International Conference, Munich, Germany, October 5-9, 2015, Proceedings, Part III 18}, \bibinfo{organization}{Springer}. pp. \bibinfo{pages}{234--241}.
\bibitem[{Roy et~al.(2023)Roy, Koehler, Ulrich, Baumgartner, Petersen, Isensee, Jaeger and Maier-Hein}]{roy2023mednext}
\bibinfo{author}{Roy, S.}, \bibinfo{author}{Koehler, G.}, \bibinfo{author}{Ulrich, C.}, \bibinfo{author}{Baumgartner, M.}, \bibinfo{author}{Petersen, J.}, \bibinfo{author}{Isensee, F.}, \bibinfo{author}{Jaeger, P.F.}, \bibinfo{author}{Maier-Hein, K.H.}, \bibinfo{year}{2023}.
\newblock \bibinfo{title}{Mednext: transformer-driven scaling of convnets for medical image segmentation}, in: \bibinfo{booktitle}{International Conference on Medical Image Computing and Computer-Assisted Intervention}, \bibinfo{organization}{Springer}. pp. \bibinfo{pages}{405--415}.
\bibitem[{Shaker et~al.(2024)Shaker, Maaz, Rasheed, Khan, Yang and Khan}]{shaker2024unetr++}
\bibinfo{author}{Shaker, A.M.}, \bibinfo{author}{Maaz, M.}, \bibinfo{author}{Rasheed, H.}, \bibinfo{author}{Khan, S.}, \bibinfo{author}{Yang, M.H.}, \bibinfo{author}{Khan, F.S.}, \bibinfo{year}{2024}.
\newblock \bibinfo{title}{Unetr++: delving into efficient and accurate 3d medical image segmentation}.
\newblock \bibinfo{journal}{IEEE Transactions on Medical Imaging} .
\bibitem[{Tolstikhin et~al.(2021)Tolstikhin, Houlsby, Kolesnikov, Beyer, Zhai, Unterthiner, Yung, Steiner, Keysers, Uszkoreit et~al.}]{tolstikhin2021mlp}
\bibinfo{author}{Tolstikhin, I.O.}, \bibinfo{author}{Houlsby, N.}, \bibinfo{author}{Kolesnikov, A.}, \bibinfo{author}{Beyer, L.}, \bibinfo{author}{Zhai, X.}, \bibinfo{author}{Unterthiner, T.}, \bibinfo{author}{Yung, J.}, \bibinfo{author}{Steiner, A.}, \bibinfo{author}{Keysers, D.}, \bibinfo{author}{Uszkoreit, J.}, et~al., \bibinfo{year}{2021}.
\newblock \bibinfo{title}{Mlp-mixer: An all-mlp architecture for vision}.
\newblock \bibinfo{journal}{Advances in neural information processing systems} \bibinfo{volume}{34}, \bibinfo{pages}{24261--24272}.
\bibitem[{Wang et~al.(2022)Wang, Xie, Lin, Iwamoto, Han, Chen and Tong}]{wang2022mixed}
\bibinfo{author}{Wang, H.}, \bibinfo{author}{Xie, S.}, \bibinfo{author}{Lin, L.}, \bibinfo{author}{Iwamoto, Y.}, \bibinfo{author}{Han, X.H.}, \bibinfo{author}{Chen, Y.W.}, \bibinfo{author}{Tong, R.}, \bibinfo{year}{2022}.
\newblock \bibinfo{title}{Mixed transformer u-net for medical image segmentation}, in: \bibinfo{booktitle}{ICASSP 2022-2022 IEEE international conference on acoustics, speech and signal processing (ICASSP)}, \bibinfo{organization}{IEEE}. pp. \bibinfo{pages}{2390--2394}.
\bibitem[{Wang et~al.(2021a)Wang, Chen, Ding, Yu, Zha and Li}]{wang2021transbts}
\bibinfo{author}{Wang, W.}, \bibinfo{author}{Chen, C.}, \bibinfo{author}{Ding, M.}, \bibinfo{author}{Yu, H.}, \bibinfo{author}{Zha, S.}, \bibinfo{author}{Li, J.}, \bibinfo{year}{2021}a.
\newblock \bibinfo{title}{Transbts: Multimodal brain tumor segmentation using transformer}, in: \bibinfo{booktitle}{Medical Image Computing and Computer Assisted Intervention--MICCAI 2021: 24th International Conference, Strasbourg, France, September 27--October 1, 2021, Proceedings, Part I 24}, \bibinfo{organization}{Springer}. pp. \bibinfo{pages}{109--119}.
\bibitem[{Wang et~al.(2021b)Wang, Xie, Li, Fan, Song, Liang, Lu, Luo and Shao}]{wang2021pyramid}
\bibinfo{author}{Wang, W.}, \bibinfo{author}{Xie, E.}, \bibinfo{author}{Li, X.}, \bibinfo{author}{Fan, D.P.}, \bibinfo{author}{Song, K.}, \bibinfo{author}{Liang, D.}, \bibinfo{author}{Lu, T.}, \bibinfo{author}{Luo, P.}, \bibinfo{author}{Shao, L.}, \bibinfo{year}{2021}b.
\newblock \bibinfo{title}{Pyramid vision transformer: A versatile backbone for dense prediction without convolutions}, in: \bibinfo{booktitle}{Proceedings of the IEEE/CVF international conference on computer vision}, pp. \bibinfo{pages}{568--578}.
\bibitem[{Wang et~al.(2024)Wang, Lin, Wu, Yu, Cheng and Yan}]{wang2024dtmformer}
\bibinfo{author}{Wang, Z.}, \bibinfo{author}{Lin, X.}, \bibinfo{author}{Wu, N.}, \bibinfo{author}{Yu, L.}, \bibinfo{author}{Cheng, K.T.}, \bibinfo{author}{Yan, Z.}, \bibinfo{year}{2024}.
\newblock \bibinfo{title}{Dtmformer: Dynamic token merging for boosting transformer-based medical image segmentation}, in: \bibinfo{booktitle}{Proceedings of the AAAI Conference on Artificial Intelligence}, pp. \bibinfo{pages}{5814--5822}.
\bibitem[{Woo et~al.(2023)Woo, Debnath, Hu, Chen, Liu, Kweon and Xie}]{woo2023convnext}
\bibinfo{author}{Woo, S.}, \bibinfo{author}{Debnath, S.}, \bibinfo{author}{Hu, R.}, \bibinfo{author}{Chen, X.}, \bibinfo{author}{Liu, Z.}, \bibinfo{author}{Kweon, I.S.}, \bibinfo{author}{Xie, S.}, \bibinfo{year}{2023}.
\newblock \bibinfo{title}{Convnext v2: Co-designing and scaling convnets with masked autoencoders}, in: \bibinfo{booktitle}{Proceedings of the IEEE/CVF Conference on Computer Vision and Pattern Recognition}, pp. \bibinfo{pages}{16133--16142}.
\bibitem[{Wu et~al.(2021)Wu, Xiao, Codella, Liu, Dai, Yuan and Zhang}]{wu2021cvt}
\bibinfo{author}{Wu, H.}, \bibinfo{author}{Xiao, B.}, \bibinfo{author}{Codella, N.}, \bibinfo{author}{Liu, M.}, \bibinfo{author}{Dai, X.}, \bibinfo{author}{Yuan, L.}, \bibinfo{author}{Zhang, L.}, \bibinfo{year}{2021}.
\newblock \bibinfo{title}{Cvt: Introducing convolutions to vision transformers}, in: \bibinfo{booktitle}{Proceedings of the IEEE/CVF international conference on computer vision}, pp. \bibinfo{pages}{22--31}.
\bibitem[{Xiao et~al.(2021)Xiao, Singh, Mintun, Darrell, Doll{\'a}r and Girshick}]{xiao2021early}
\bibinfo{author}{Xiao, T.}, \bibinfo{author}{Singh, M.}, \bibinfo{author}{Mintun, E.}, \bibinfo{author}{Darrell, T.}, \bibinfo{author}{Doll{\'a}r, P.}, \bibinfo{author}{Girshick, R.}, \bibinfo{year}{2021}.
\newblock \bibinfo{title}{Early convolutions help transformers see better}.
\newblock \bibinfo{journal}{Advances in neural information processing systems} \bibinfo{volume}{34}, \bibinfo{pages}{30392--30400}.
\bibitem[{Xie et~al.(2021a)Xie, Wang, Yu, Anandkumar, Alvarez and Luo}]{xie2021segformer}
\bibinfo{author}{Xie, E.}, \bibinfo{author}{Wang, W.}, \bibinfo{author}{Yu, Z.}, \bibinfo{author}{Anandkumar, A.}, \bibinfo{author}{Alvarez, J.M.}, \bibinfo{author}{Luo, P.}, \bibinfo{year}{2021}a.
\newblock \bibinfo{title}{Segformer: Simple and efficient design for semantic segmentation with transformers}.
\newblock \bibinfo{journal}{Advances in neural information processing systems} \bibinfo{volume}{34}, \bibinfo{pages}{12077--12090}.
\bibitem[{Xie et~al.(2021b)Xie, Zhang, Shen and Xia}]{xie2021cotr}
\bibinfo{author}{Xie, Y.}, \bibinfo{author}{Zhang, J.}, \bibinfo{author}{Shen, C.}, \bibinfo{author}{Xia, Y.}, \bibinfo{year}{2021}b.
\newblock \bibinfo{title}{Cotr: Efficiently bridging cnn and transformer for 3d medical image segmentation}, in: \bibinfo{booktitle}{Medical Image Computing and Computer Assisted Intervention--MICCAI 2021: 24th International Conference, Strasbourg, France, September 27--October 1, 2021, Proceedings, Part III 24}, \bibinfo{organization}{Springer}. pp. \bibinfo{pages}{171--180}.
\bibitem[{Yang et~al.(2023)Yang, Zhang, Chen, Jiang, Feng, Pu and Sun}]{yang2023ucunet}
\bibinfo{author}{Yang, S.}, \bibinfo{author}{Zhang, X.}, \bibinfo{author}{Chen, Y.}, \bibinfo{author}{Jiang, Y.}, \bibinfo{author}{Feng, Q.}, \bibinfo{author}{Pu, L.}, \bibinfo{author}{Sun, F.}, \bibinfo{year}{2023}.
\newblock \bibinfo{title}{Ucunet: A lightweight and precise medical image segmentation network based on efficient large kernel u-shaped convolutional module design}.
\newblock \bibinfo{journal}{Knowledge-Based Systems} \bibinfo{volume}{278}, \bibinfo{pages}{110868}.
\bibitem[{Zhou et~al.(2023)Zhou, Guo, Zhang, Han, Yu, Wang and Yu}]{zhou2023nnformer}
\bibinfo{author}{Zhou, H.Y.}, \bibinfo{author}{Guo, J.}, \bibinfo{author}{Zhang, Y.}, \bibinfo{author}{Han, X.}, \bibinfo{author}{Yu, L.}, \bibinfo{author}{Wang, L.}, \bibinfo{author}{Yu, Y.}, \bibinfo{year}{2023}.
\newblock \bibinfo{title}{nnformer: Volumetric medical image segmentation via a 3d transformer}.
\newblock \bibinfo{journal}{IEEE Transactions on Image Processing} .
\bibitem[{Zhou et~al.(2018)Zhou, Rahman~Siddiquee, Tajbakhsh and Liang}]{zhou2018unet++}
\bibinfo{author}{Zhou, Z.}, \bibinfo{author}{Rahman~Siddiquee, M.M.}, \bibinfo{author}{Tajbakhsh, N.}, \bibinfo{author}{Liang, J.}, \bibinfo{year}{2018}.
\newblock \bibinfo{title}{Unet++: A nested u-net architecture for medical image segmentation}, in: \bibinfo{booktitle}{Deep Learning in Medical Image Analysis and Multimodal Learning for Clinical Decision Support: 4th International Workshop, DLMIA 2018, and 8th International Workshop, ML-CDS 2018, Held in Conjunction with MICCAI 2018, Granada, Spain, September 20, 2018, Proceedings 4}, \bibinfo{organization}{Springer}. pp. \bibinfo{pages}{3--11}.

\end{thebibliography}



\end{document}